\documentclass{ws-mpla}

\usepackage{tikz}
\usetikzlibrary{decorations.pathmorphing,decorations.markings}

\usepackage{hyperref}  
\usepackage{graphicx}  
\usepackage{pdfpages}  
\usepackage{rotating}  
\usepackage{slashed}   

\def\epem{\ensuremath{e^+e^-}}
\def\beq{\begin{equation}}
\def\eeq{\end{equation}}
\def\beqa{\begin{eqnarray}}
\def\eeqa{\end{eqnarray}}

\def\BtoDsttaunuGen{\ensuremath{B\to \bar D^{(*)} \tau^+ \nu_\tau}}

\def\BtoDsttaunu{\ensuremath{B\to \bar D^{*} \tau^+ \nu_\tau}}
\def\BztoDsttaunu{\ensuremath{B^0\to D^{*-} \tau^+ \nu_\tau}}
\def\BztoDtaunu{\ensuremath{B^0\to D^{-} \tau^+ \nu_\tau}}
\def\BptoDsttaunu{\ensuremath{B^+\to \bar D^{*0} \tau^+ \nu_\tau}}
\def\BptoDtaunu{\ensuremath{B^+\to \bar D^{0} \tau^+ \nu_\tau}}

\def\BtoDtaunu{\ensuremath{B\to \bar D \tau^+ \nu_\tau}}

\def\BtoDstlnuGen{\ensuremath{B\to \bar D^{(*)} \ell^+ \nu_\ell}}
\def\BtoDstlnu{\ensuremath{B\to \bar D^* \ell^+ \nu_\ell}}

\def\Btotaunu{\ensuremath{B^+\to \tau^+ \nu_\tau}}

\def\gev{\ensuremath{\rm GeV}}
\def\mev{\ensuremath{\rm MeV}}
\def\fours{\ensuremath{\Upsilon(4S)}}

\def\babar{\mbox{\slshape B\kern-0.1em{\small A}\kern-0.1em
    B\kern-0.1em{\small A\kern-0.2em R}}}
\def\babartbl{\mbox{\slshape B\kern-0.1em{\scriptsize A}\kern-0.1em
    B\kern-0.1em{\scriptsize A\kern-0.2em R}}}

\def\sqrts{\ensuremath{\sqrt{s}}}
\def\btag{\ensuremath{B_{\rm tag}}}
\def\bsig{\ensuremath{B_{\rm sig}}}
\def\br{\ensuremath{{\cal B}}}

\def\mes{\ensuremath{m_{\rm ES}}}
\def\DE{\ensuremath{\Delta E}}
\def\Eex{\ensuremath{E_{\rm extra}}}
\def\mmisssq{\ensuremath{m^2_{\rm miss}}}

\def\pmiss{\ensuremath{p_{\rm miss}}}

\def\RDstGen{\ensuremath{R(D^{(*)})}}
\def\RD{\ensuremath{R(D)}}
\def\RDst{\ensuremath{R(D^*)}}

\def\KS{\ensuremath{K_S^0}}
\def\KL{\ensuremath{K_L^0}}

\begin{document}


\catchline{}{}{}{}{}

\title{\boldmath $B$-MESON DECAYS INTO FINAL STATES WITH A $\tau$ LEPTON}

\author{ABNER SOFFER}

\address{School of Physics and Astronomy, Tel Aviv University\\
Tel Aviv, 69978, Israel\\
asoffer@tau.ac.il}

\maketitle

\pub{Received (Day Month Year)}{Revised (Day Month Year)}

\begin{abstract}

Decays of $B$ mesons into final states containing a $\tau$ lepton are
sensitive to new charged-current interactions that break lepton-flavor
universality. These decays have been studied only at $\epem$
colliders, where the low-background environment and well-known initial
state make it possible to observe small signals with undetectable
neutrinos.  In particular, the large data samples of the $B$~factories
and recent advances in techniques for full-event reconstruction have
led to evidence for the decay \Btotaunu\ and unambiguous observation
of the decays \BtoDsttaunuGen. These results exclude
large regions of the parameter space for a variety of new-physics
models.  Furthermore, the branching fraction for \BtoDsttaunuGen\ has
been measured to be higher than the standard-model expectation by 
more than $3$ standard deviations, making this an interesting topic for
further research.
This letter reviews the theoretical and experimental status of this
topic, summarizing the results at this time and outlining the 
path for further improvements.

\end{abstract}

\ccode{PACS Nos.: 14.40.Nd, 14.60.Fg, 13.20.He, 13.20.-v}

\section{Introduction}	
\label{sec:intro}

The decays \BtoDsttaunuGen\ and \Btotaunu\ are well suited for
searching for effects of new physics (NP) in charged-current
interactions. In particular, the presence of third-generation
fermions in both the initial and final-state leads to sensitivity to new
particles that couple more strongly to heavy fermions, such as a
charged Higgs.

The multiple neutrinos produced in these exclusive decays make it impossible to
reconstruct the invariant mass of the $B$ meson and use it for 
background rejection. Therefore, their study 
requires use of additional constraints related to the
production of the $B$ meson. Such constraints are available at $B$
factories, which collide electrons and positrons at an average
center-of-mass energy of $\sqrts\approx 10.58~\gev$, corresponding to
the mass $m_{\fours}$ of the \fours\ resonance. As a result, the $B$
factories \babar\cite{TheBABAR:2013jta,Aubert:2001tu} and
Belle\cite{Abashian:2000cg} have provided the only measurements of
these decays.

The $B$-factory results include evidence for \Btotaunu\ and more than
$3.4$-standard-deviation ($\sigma$) difference between the \BtoDsttaunuGen\
decay rates and the expectation of the standard model (SM).  Better
understanding of this tension will come from improved measurements
of the decay rates and the angular distributions at the current
$B$~factories. During the next decade, the
Belle-II\cite{Abe:2010sj} experiment, which will have an integrated
luminosity over 30 times greater than that of the combined \babar\ and Belle
datasets, will provide accurate measurements that should pinpoint
possible NP contributions to these decays with great precision.

This paper is organized as follows. In Sec.~\ref{sec:theory} we
discuss the theoretical background and predictions for measurements of
\BtoDsttaunuGen\ and \Btotaunu. Sec.~\ref{sec:btag} outlines the
experimental technique of full-event reconstruction, which is unique
to the $B$~factories and critical for enabling the study of these
decays.
We review the experimental results in Sec.~\ref{sec:results}, and
discuss the implications for new physics in
Sec.~\ref{sec:interp}. Concluding remarks and the outlook for 
future measurements are given in Sec.~\ref{sec:conc}.

\section{Theory and Predictions}
\label{sec:theory}

\begin{figure}[!btph]
\centerline{


\input diag-defs.tex

\begin{tabular}{cc}
\begin{tikzpicture}[thick,scale=1.0]
\draw (0,\yqbar-0.5) node[]{(a)};
\draw[antiparticle] (\xleft,1) node[black,left]{$\bar b$} -- (0,1);
\draw[antiparticle] (0,1) -- (\xright,1) node[black,right]{$\bar c$};
\draw[particle] (\xleft,\yqbar) node[black,left] {$q$} -- (\xright,\yqbar);
\draw[boson] (0,1) -- (0.5,1.75);
\draw[antiparticle] (\xvtx,1.75) -- (\xright,1.5) node[black,right] {$\tau^+$} ;
\draw[particle] (\xvtx,1.75) -- (\xright,2) node[black,right] {$\nu_\tau$};
\end{tikzpicture}

& 
\hspace{0.5cm}

\begin{tikzpicture}[thick,scale=1.0]
\draw (0,-0.8) node[]{(b)};
\draw[antiparticle] (-1.2,0.5) node[black,left]{$\bar b$} -- (-0.5,0);
\draw[particle] (-1.2,-0.5) node[black,left] {$u$} -- (-0.5,0);
\draw[boson] (-0.5,0) -- (0.5,0);
\draw[antiparticle] (0.5,0) -- (1.2, 0.5) node[black,right] {$\tau^+$} ;
\draw[particle] (0.5,0) -- (1.2, -0.5) node[black,right] {$\nu_\tau$};
\end{tikzpicture}

\end{tabular}

}
\vspace*{8pt}
\caption{Standard-model Feynman diagrams for \BtoDsttaunuGen\ (a)
and \Btotaunu\ (b). 
\protect\label{fig:diag-Dtaunu-taunu}}
\end{figure}
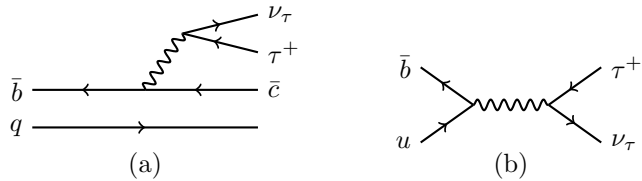

\subsection{\BtoDsttaunuGen\ Theory}
\label{sec:th-Dtaunu}

The SM Feynman diagram for \BtoDsttaunuGen\ is shown in
Fig.~\ref{fig:diag-Dtaunu-taunu}(a).  The decay takes place via $W$
emission, and in this respect is identical to \BtoDstlnuGen\ (where
we use $\ell$ to indicate an electron or muon).  However, 
\BtoDsttaunuGen\ is also sensitive
to NP that preferentially impacts heavy fermions and thus escapes
detection in \BtoDstlnuGen. A widely discussed example is mediation by
a charged Higgs boson.
The effective Hamiltonian that accounts for the SM plus new vector,
scalar, and tensor interactions
is\cite{Faller:2011nj,Datta:2012qk,Tanaka:2012nw}
\beqa
{\cal H}_{\rm eff} = && {4 G_F V_{cb} \over \sqrt{2}} 
\bigl[
(1 + V_L) \left(\bar c \gamma_\mu P_L b\right)
          \left(\bar\tau\gamma^\mu P_L \nu_\tau\right)
  + V_R \left(\bar c \gamma_\mu P_R b\right) 
        \left(\bar\tau \gamma^\mu P_L \nu_\tau\right)
\nonumber\\
 && + S_L \left(\bar c P_L b\right) 
          \left(\bar \tau P_L \nu_\tau\right) 
    + S_R \left(\bar c P_R b\right) 
          \left(\bar \tau P_L \nu_\tau\right) 
\nonumber\\
 && + T_L \left(\bar c \sigma^{\mu\nu} P_L b\right) 
          \left(\bar \tau \sigma_{\mu\nu} P_L \nu_\tau\right) 
       \bigr] + H.c.,
\label{eq:dtaunu-H}
\eeqa
where $G_F$ is the Fermi coupling constant, $V_{cb}$ 
is the Cabibbo-Kobayashi-Maskawa (CKM) matrix
elements\cite{Cabibbo:1963yz,Kobayashi:1973fv}, 
$\gamma_\mu$ are the Dirac matrices, 
$\sigma_{\mu\nu} = i[\gamma_\mu, \gamma_\nu]/2$,
$P_{L,R} \equiv (1 \mp
\gamma_5)/2$ are the left and right projection operators, 
and $V_{L,R}$, $S_{L,R}$, and $T_L$ are complex Wilson coefficients
that govern the NP 
contributions\footnote{Eq.~(\ref{eq:dtaunu-H}) ignores the possibility of
  lepton-flavor violation in the leptonic terms\cite{Tanaka:2012nw},
  since it is unobservable in this measurement.}.
The SM corresponds to $V_{L,R} = S_{L,R} = T_L = 0$.

In what follows we take $V_{L,R} = T_L = 0$ and focus on the scalar
terms in Eq.~(\ref{eq:dtaunu-H}) and on their implications for a
charged Higgs boson. These terms describe the most general
two-Higgs-doublet model, also known as type-III 2HDM.  The more
restricted type-II 2HDM, which is the Higgs sector of the minimal
supersymmetric standard model, corresponds to $S_L=0, S_R = -m_b
m_\tau \tan^2\beta / m^2_{H^\pm}$, where $\tan^2\beta$ is the ratio
between the vacuum expectation values of the two Higgs doublets and
$m_{H^\pm}$ is the mass of the charged Higgs.
The differential decay rate is then given by\cite{ref:dtaunu-theory-fajfer}
\beqa
{d\Gamma \over dq^2} = &&
    {G_F^2 |V_{cb}|^2 p^*_{D^{(*)}}q^2 \over 96 \pi^3 m_B^2}
    \left(1 - {m_\tau^2 \over q^2}\right)^2  \nonumber\\
    && \left[\left(|H_+|^2 + |H_-|^2 + |H_0|^2\right) 
        \left(1 + {m_\tau^2 \over 2q^2}\right) 
    + {3 \over 2} {m_\tau^2 \over q^2} |H_s|^2\right],
\label{eq:Dtaunu-rate}
\eeqa
where $p^*_{D^{(*)}}$ is the momentum of the $D^{(*)}$ in the 
$B$-meson rest frame, 
$q^2$ is the squared four momentum of the leptons,
and $H_x$ are $q^2$-dependent helicity amplitudes.
The scalar terms in Eq.~(\ref{eq:dtaunu-H}) affect only the $H_s$
amplitude\cite{ref:dtaunu-theory-fajfer,Tanaka:1994ay,Kamenik:2008tj}:
\beq
H_s = H_s^{\rm SM} \left[1 + (S_R \pm S_L) 
                               {q^2 \over m_\tau(m_b \mp m_c)} \right],
\label{eq:HsNP}
\eeq
where the upper sign is for \BtoDtaunu\ and the lower is for \BtoDsttaunu.

Hadronic uncertainties associated with the form factors that
govern the helicity amplitudes are reduced, and the
uncertainties due to constants such as $V_{cb}$ and $G_F$ are eliminated,
when one studies the ratios of decay rates
\beq
\RDstGen \equiv {\Gamma(\BtoDsttaunuGen)\over \Gamma(\BtoDstlnuGen)}.
\label{eq:R-def}
\eeq
The numerator is obtained by integrating Eq.~(\ref{eq:Dtaunu-rate}),
and the denominator comes from the same expression with the
replacement of $m_\tau$ by $m_\ell$. The different $q^2$ spectra of
the two processes are accounted for in the helicity amplitudes.
The SM values of these ratios have been
calculated\cite{ref:dtaunu-theory-fajfer,ref:bbr-dtaunu-2012,ref:bbr-dtaunu-2013}
using form factors obtained from
\BtoDstlnu\ decays\cite{ref:hfag-FF} and heavy quark effective
theory\cite{Tanaka:2010se}:
\beqa
R_{\rm SM}(D)   &=& 0.297 \pm 0.017, \nonumber\\
R_{\rm SM}(D^*) &=& 0.252 \pm 0.03.
\label{eq:R-SM}
\eeqa
An unquenched lattice-QCD calculation\cite{Bailey:2012jg} yields 
for $R_{\rm SM}(D)$ a higher yet consistent value:
\beq
R_{\rm SM}(D) = 0.316 \pm 0.012 \pm 0.07,
\label{eq:R-SM-lattice}
\eeq
where here and throughout the article, the first set of uncertainties
is statistical and the second is systematic. 
A similar prediction,
\beq
R_{\rm SM}(D) = 0.31 \pm 0.02,
\label{eq:R-SM-mintheory}
\eeq
has been obtained with only minimal reliance on theoretical
input\cite{Becirevic:2012jf}.
Eq.~(\ref{eq:R-SM}) and
measurements\cite{ref:hfag-2009,Nakamura:2010zzi,Dungel:2010uk} of
$\br(\BtoDstlnuGen)$ yield the expected branching
fractions\cite{ref:dtaunu-theory-fajfer}
\beqa
\br(\BptoDtaunu)_{\rm SM} &=& (0.66 \pm 0.05)\% \nonumber\\
\br(\BztoDtaunu)_{\rm SM} &=& (0.64 \pm 0.05)\% \nonumber\\
\br(\BptoDsttaunu)_{\rm SM} &=& (1.43 \pm 0.05)\% \nonumber\\
\br(\BztoDsttaunu)_{\rm SM} &=& (1.29 \pm 0.06)\%.
\label{eq:Dtaunu-brs}
\eeqa

Eqs.~(\ref{eq:Dtaunu-rate}) and~(\ref{eq:HsNP}) give the impact of the
NP terms on the rate ratios,
\beqa
\RD &=& R_{\rm SM}(D) + A'_D \Re (S_R + S_L) + B'_D (S_R + S_L|^2,
      \nonumber\\
\RDst &=& R_{\rm SM}(D^*) + A'_{D^*} \Re (S_R - S_L) + B'_{D^*} (S_R - S_L|^2,
\label{eq:R-NP-III}
\eeqa
where $A'_{D^{(*)}}$ and $B'_{D^{(*)}}$ are coefficients that depend on the
form factors and the quark masses.
In a type-II 2HDM, this becomes
\beq
\RDstGen_{\rm type\ II} = R_{\rm SM}(D^{(*)}) 
                                 + A_{D^{(*)}} {\tan^2\beta \over m^2_{H^\pm}}
                                 + B_{D^{(*)}} {\tan^4\beta \over m^4_{H^\pm}}.
\label{eq:R-NP-II}
\eeq
The coefficients in these expressions have been
calculated\cite{ref:bbr-dtaunu-2013} to be
\beqa
A_D       = -3.25 \pm 0.32~\gev^{2} &,& \ \ 
  A_{D^*} = -0.230 \pm 0.029~\gev^{2} , \nonumber\\
  B_D     = 16.9 \pm 2.0~\gev^{4} &,& \ \ 
  B_{D^*} = 0.643 \pm 0.085~\gev^{4} , \nonumber\\
A'_{D^*} = -{A_{D^*} \over m_\tau m_b} &,& \ \ 
  B'_{D^*} =  {B_{D^*} \over m_\tau^2 m_b^2}.
\label{eq:AB-coeffs}
\eeqa

In addition to the total branching fraction and the $q^2$ dependence
of the decay rate, angular distributions can also be used to study NP
contributions, as can CP-violating triple-product asymmetries that are
non-zero when NP couplings are complex. The impact of NP contributions
on the angular differential decay rates has been evaluated
theoretically\cite{Tanaka:2010se,Datta:2012qk,Duraisamy:2013pia}, 
but has not yet been studied
experimentally.

\subsection{\Btotaunu\ Theory}

The SM Feynman diagram for \Btotaunu\ is shown in
Fig.~\ref{fig:diag-Dtaunu-taunu}(b).  Eq.~(\ref{eq:dtaunu-H})
describes the effective Lagrangian for this process, following the
quark replacement $c \to u$ and accounting for a possible flavor
dependence of the couplings.  We again take $V_{L,R} = T_L = 0$ to
obtain the branching fraction prediction for the SM plus a new scalar
interaction\cite{Crivellin:2012ye},
\beqa
\br(\Btotaunu) = && {G_F^2 m_B m_\tau^2 \over 8\pi}
               \left(1 - {m_\tau^2 \over m_B^2} \right)^2
	       f_B^2 |V_{ub}|^2 \tau_B  \nonumber\\
           &&\times \left|1 + {m_B^2 \over m_\tau m_b} (S_R - S_L)\right|^2,
\label{eq:br-taunu}
\eeqa
where $f_B = 189 \pm 4~\mev$ is the $B$-meson decay
constant\cite{Na:2012kp}.

The largest uncertainty on the SM-predicted value of this branching
fraction arises from the CKM element $|V_{ub}|$.  The Particle Data
Group\cite{pdg12} has calculated the world average value $|V_{ub}|=(4.15 \pm
0.49)\times 10^{-3}$, after scaling the measurement uncertainties by a
factor of $2.6$ to account for the roughly $3\sigma$
difference\cite{pdg12,ref:hfag} between the value obtained from the
inclusive branching fraction $\br(B\to X_u \ell^+ \nu_\ell)$ and the one
from the exclusive branching fraction $\br(B\to \pi \ell^+ \nu_\ell)$.
Eq.~(\ref{eq:br-taunu}) then leads to the prediction
\beq
\br_{\rm SM}(\Btotaunu) = (1.23 \pm 0.29)\times 10^{-4};
\label{eq:taunu-SM-PDG}
\eeq
The $|V_{ub}|$ values obtained from global unitarity-triangle fits
performed by the CKMfitter\cite{ref:ckmfitter} and
UTfit\cite{ref:utfit} collaborations favor the
$\br(B\to \pi \ell^+ \nu_\ell)$ results. The
$\br(\Btotaunu)$ values predicted by these fits are
\beq
\br_{\rm SM}(\Btotaunu) = \left\{
   \begin{matrix}
     0.739 ^{+0.090} _{-0.070} \times 10^{-4}  & {\rm CKMfitter} \\[10pt]
     0.81 \pm 0.07)\times 10^{-4}  & {\rm UTfit} \\
   \end{matrix}
  \right. .
  \label{eq:taunu-SM-fits}
\eeq
The dependence on $V_{ub}$ cancels in the ratio of branching
fractions\cite{Khodjamirian:2011ub,Crivellin:2009sd,Buras:2010pz,Lunghi:2010gv}
\beq
R' = {\tau_{B^0} \over \tau_{B^+}} 
   {\br(\Btotaunu) \over \br(B^0\to \pi^-\ell ^+\nu_\ell)},
\label{eq:R'}
\eeq
which is $R'=0.31 \pm 0.06$ in the
SM\cite{Fajfer:2012jt}.

\section{The Technique of Full-Event Reconstruction}
\label{sec:btag}

A $B$~factory is a high-luminosity \epem\ collider with an average
center-of-mass (CM) collision energy \sqrts\ that equals the
\fours\ mass\cite{pdg12}, $m_\fours = 10.5794 \pm
0.0012~\gev$\footnote{We ignore the ${\cal O}(\mev)$ impact of
  initial-state radiation, which is anyway calibrated out in the
  measurement of \sqrts.}  In what follows, we take all kinematic
quantities in the average CM frame.
The \fours\ decays promptly to two $B$ mesons, so that the $B$ energy
equals $\sqrts/2$ to within half the collision-energy
spread, which is $\sigma_{\sqrts} \approx 5~\mev$ at the current
$B$~factories\cite{Aubert:2001tu}.
The momenta of the two $B$ mesons, which average $330~\mev$, are equal
to within $\sigma_{\sqrts}$ and opposite to
within $2^\circ$.

These event characteristics are used to address the difficulties
caused by undetectable neutrinos in rare $B$-meson decays.
This is done by reconstructing not only the signal $B$ decay of
interest (labeled \bsig), but also the other $B$ meson in the event,
known as the tag~$B$ (labeled \btag). 
In such full-event reconstruction, it is typically required that all
charged-particle tracks be assigned to one of the two $B$
candidates. Furthermore, the energy \Eex\ of unassigned calorimeter
clusters or photon candidates is required to be low, typically around
$1~\gev$. This requirement reflects the fact that such ``extra''
energy arises not only from missing particles in background events,
but also from calorimeter noise, previous events, and scattered
particles from the interaction of hadrons with the calorimeter
material in signal events.

By attempting to account for the origin of all particles in the event,
full-event reconstruction reduces the rate of the combinatorial
background, which arises from random combinations of particles that
happen to satisfy the selection criteria.
Furthermore, if the tag~$B$ is fully and
correctly reconstructed in a hadronic final state, the kinematic constraints
described above yield a measurement of the 4-momentum of the missing
neutrinos, further aiding with signal identification
and enabling the calculation of quantities in the signal-$B$ rest frame.

The disadvantage of full-event reconstruction is the low efficiency
for reconstructing the large number of particles produced in a typical
tag-$B$ decay\cite{Brandenburg:1999cs}. Tag-$B$ final states with a high
multiplicity of charged tracks and $\pi^0$ mesons tend to also have
low purity, defined as the fraction of correctly reconstructed decays
among all selected \btag\ candidates, due to the high combinatorial
background.
Nevertheless, the large datasets of \babar\ and Belle and increasing
sophistication in the application of \btag-reconstruction
techniques have made this technique an indispensable tool for the study
of rare $B$ decays and decays with multiple neutrinos.

Tag-$B$ reconstruction is performed by one of three techniques:
hadronic tagging, semileptonic tagging, or inclusive tagging,
depending on the \btag\ final state and
reconstruction method. The methods are generally complementary,
with each having different advantages,
disadvantages, and relative importance that depends, among other
factors, on the signal-$B$ decay of interest. 
The details of each of each of these techniques are described in the
following subsections.

\subsection{Hadronic Tagging}
\label{sec:hadronic-tag}

In hadronic tagging, the tag $B$ is fully reconstructed from its decay into
a hadronic final state. 
Use of this technique was first reported by the ARGUS
collaboration\cite{Albrecht:1990wa}.
Since hadronic tagging provides the 4-momentum $p^\mu_{\rm tag}$ of
the tag~$B$, one can calculate the four-momentum of the
undetectable neutrinos and their invariant mass, known as
the missing mass:
\beqa
\pmiss^\mu &=& \left<p^\mu_{\epem}\right> - p^\mu_{\rm tag} - p^\mu_{Y},
\nonumber\\
\mmisssq &=& \pmiss^2,
\label{eq:pmiss-generic}
\eeqa
where $Y$ denotes the visible particles in the final state of the
signal decay and $\left<p^\mu_{\epem}\right>$ is
the average $\epem$ four-momentum, which is measured from the calibrated 
accelerator-beam parameters. 
The missing mass is useful for signal-background separation.
Furthermore, the well-defined rest frame of the signal~$B$ allows
calculation of $q^2$ and $p^*_{D^{(*)}}$ of Eq.(\ref{eq:Dtaunu-rate}),
and additional variables that can be used for background suppression.

The tag $B$ is reconstructed from decays that proceed via 
$b\to c \bar u d$ or $b\to c \bar c s$ transitions,
which have the largest branching
fractions due to their large CKM matrix elements. 
In most $B$ decays, the charm quarks hadronize into a
charmed or charmonium meson. This is utilized for
combinatoric-background reduction by reconstructing a $D^{(*)+}$,
$D^{(*)0}$, $D_s^{(*)-}$, or $J/\psi$ candidate, which is selected
based on an invariant-mass criterion.

The kinematic characteristics of $\fours \to
B\bar B$ events are brought into play by the use of two standard
variables,
\beq
\DE = E_{\rm tag} - \sqrts/2, \ \ \ \mes = \sqrt{s/4 - p^2_{\rm tag}},
\label{eq:mes-de}
\eeq
where $E_{\rm tag}$ and $p_{\rm tag}$ are, respectively, the
reconstructed energy and momentum of the \btag\ candidate. The
expression for \mes\ is essentially the \btag\ invariant mass,
with $E_{\rm tag}$ replaced by $\sqrts/2$, which is much better
known and measured independently of $p_{\rm tag}$.
For correctly reconstructed \btag\ candidates, $\DE$ and \mes\ have
nearly normal distributions that peak at $0$ and $m_B$, with 
typical widths of 
$10-35~\mev$ and $\sigma_{\sqrts}/2$, respectively. 
The typical background distribution under the signal
peak is approximately linear in $\DE$ and rapidly falling in
\mes\ with the diminishing phase space.  Basic \btag\ selection is
accomplished by requiring $\DE$ and \mes\ to be within
mode-dependent distances of their peak values.

In order to maximize efficiency and purity, final states with low 
particle multiplicity are preferred. However, the small total 
branching fraction of such decays necessitates use of 
higher-multiplicity decays as well.
The number of \btag\ reconstruction modes and the ways these modes are
selected and handled have evolved over time. In the most recent
hadronic-tagging analyses, \babar\ and Belle reconstructed well over
1000 \btag\ modes, leading to approximately a four-fold increase in
the effective \btag\ efficiency relative to the earliest $B$-factory
hadronic-tagging analysis\cite{Aubert:2003yh}. This necessarily
introduced many low-purity \btag\ decays, requiring removal of as many
incorrectly reconstructed \btag\ candidates as possible while still
maintaining high \btag-reconstruction efficiency. 
The two collaborations developed different approaches for
carrying this out.

The approach taken by \babar\ was to simply remove the lowest-purity
\btag\ modes, where the purity of each mode was determined in a
\bsig-decay-specific way from simulated events containing a true
\bsig\ decay and a generic \btag\ decay.  This took advantage
of the dependence of the purity on \bsig-specific factors, such as
final-state multiplicity.

Belle considered the \btag\ decay separately from the signal-$B$
decay, but applied a more sophisticated method of using
\btag\ information to obtain high purity and efficiency\cite{ref:bll-hadtag}.
Tag-$B$ reconstruction was divided into four stages: (1) tracks,
photons, $\KS $, and $\pi^0$ candidates; (2) charmed-meson candidates;
(3) excited charmed-meson candidates; and (4) $B$ candidates.  At each
stage, neural-network algorithms were used to determine the
probability that the \btag\ components were correctly reconstructed,
using input variables relevant for that stage. 
%
The product of the neural-network outputs of each stage was also used
as an input variable for the subsequent stage. The output of the final
neural-network was used, along with \mes\ and $\DE$, for final 
\btag-candidate selection\cite{ref:bll-taunu-2013}.
%

It is interesting to consider the possibility of further improvements
in the purity and efficiency of hadronic tagging. The \babar\ method
is better at exploiting the signal-$B$ decay, and the Belle method
makes better use of information within the tag-$B$ decay.  
Combining the two approaches by executing the Belle method for each
signal-$B$ mode separately may lead to further improvements.

\subsection{Semileptonic Tagging}
In semileptonic tagging, the tag $B$ is reconstructed in the
four semileptonic final states $D^{(*)}\ell^-\nu_\ell$, which
make up $(7.92 \pm 0.17)\%$ and $(7.11 \pm 0.22)\%$ of the 
$B^-$ and $\bar B^0$ branching fractions, respectively\cite{pdg12}.
So far, only the most favorable
charmed-meson final states 
$D^0 \to K^-\pi^+$, $K^-\pi^+\pi^0$, $K^-\pi^+\pi^-\pi^+$,
$\KS \pi^+\pi^-$, and $D^+ \to K^-\pi^+\pi^+$,
$\KS \pi^+$ have been used.
The $D^*$ decays have been $D^{*+}\to D^0\pi^+$, $D^{*+}\to
D^+\pi^0$, $D^{*0}\to D^0\pi^0$, and $D^{*0}\to D^0\gamma$.
In some cases\cite{Aubert:2007bx}, there was no attempt to
reconstruct the soft $\pi^0$ or photon from the $D^{*0}$ decay, in
order to increase efficiency and also accept $D^{**0}\to
D^0\pi^0$ decays, at the cost of increased background.

Although the \btag\ is not fully reconstructed,
four-momentum conservation in its decay and the fact that its
CM-frame 4-momentum is known 
yield the CM-frame angle between the momentum vector
of the \btag\ and that of the $D^{(*)}\ell$ system,
\beq
\cos\theta_{B-D^{(*)}\ell} = {E_{D^{(*)}\ell} \sqrts  - 
      m_B^2 - m_{D^{()*}\ell}^2
             \over 2 p_{D^{(*)}\ell} \sqrt{s/4 - m_B^2} },
\label{eq:cosThetaBY}
\eeq
where $E_{D^{(*)}\ell}$, $m_{D^{(*)}\ell}$, and $p_{D^{(*)}\ell}$ are,
respectively, the energy, invariant mass, and 3-momentum of the
$D^{(*)}\ell$ system.
Tag-$B$ candidates are required to have $\cos\theta_{B-D^{*}\ell}$ in
a range somewhat larger than $[-1,1]$, to allow for detector
resolution and for the loss of a soft pion or final-state-radiation
photons from an otherwise correctly reconstructed
\btag\ candidate. Background
candidates may have $\cos\theta_{B-D^{*}\ell}$ values well beyond the selection
range.

\subsection{Inclusive Tagging}

In the inclusive-tagging method, one reconstructs the signal-$B$
candidate and then attempts to reconstructs the \btag\ from all
remaining tracks and photon candidates, while making no attempt to
break the \btag\ decay down according to known decay channels of the
$B$ meson. 
In further contrast to the hadronic-tagging method, only loose
requirements on \mes\ and $\DE$ are applied, to allow for some lost
particles, in particular $\KL$ mesons, which are produced copiously in charm
and bottom decays.
All this makes inclusive tagging simpler and more efficient
than hadronic tagging, while providing less background rejection.
This technique was first used by the CLEO collaboration in the 
measurements of the $B^0\to \pi^- \ell^+\nu_\ell$ and $B^0\to \rho^-
\ell^+\nu_\ell$ branching fractions\cite{Alexander:1996qu}.

\section{Experimental Results}
\label{sec:results}

We describe the results for \BtoDsttaunuGen\ in Sec.~\ref{sec:Dtaunu},
and those for \Btotaunu\ in Sec.~\ref{sec:taunu}. The discussion focuses
on the latest and most precise measurements, summarizing older results 
briefly.

In addition to the full-event-reconstruction variables introduced in
Sec.~\ref{sec:btag}, each data analysis used requirements on various
kinematic variables to suppress the background. Some of these
variables quantified the difference between the isotropic distribution
of particle momenta in $\fours \to B\bar B$ events and the jet-like
structure of ``continuum'' $\epem\to q\bar q$ events, where $q$
represents a $u$, $d$, $s$, or $c$ quark. The other variables were
analysis-specific, and were related to the degree of missing energy and
momentum in the multi-neutrino signals, angular correlations between
particle momenta, or invariant masses of intermediate resonances. The
description here glosses over such details, focusing on the main
measurement techniqnes and results.

\subsection{\BtoDsttaunuGen\ Results}
\label{sec:Dtaunu}

We describe here the four \babar\ and Belle journal publications on
\BtoDsttaunuGen, as well as a preliminary Belle result that has
been used in a combination of the different measurements,
which is presented in Sec.\ref{sec:dtaunu-results-summary}. 
Physics interpretations of the results are discussed in Sec.~\ref{sec:interp}.

\subsubsection{Belle Inclusive-Tagging Measurement}
\label{sec:dtaunu-belle-inctag}

Belle made the first observation of a \BtoDsttaunuGen\ decay in 2007,
using a data sample of $535\times 10^6$ $B\bar B$ pairs and the
inclusive-tagging method\cite{Matyja:2007kt}. The signal $B$ was
reconstructed in the decay \BztoDsttaunu, taking advantage of
the efficient background suppression provided by $D^{*-}\to\bar D^0 \pi^-$
reconstruction. The $\bar D^0$ was reconstructed only in the channels
$\bar D^0\to K^+\pi^-$ and $\bar D^0\to K^+\pi^-\pi^0$, and the
$\tau^+$ was reconstructed in $\tau^+ \to e^+ \nu_e \bar\nu_\tau$ and
$\tau^+ \to \pi^+ \bar\nu_\tau$, the latter channel being also
sensitive to $\tau^+ \to \rho^+ \bar\nu_\tau$.

\begin{figure}[htbp]
\addtocounter{footnote}{-1}
\centerline{
\includegraphics[width=1.0\textwidth]{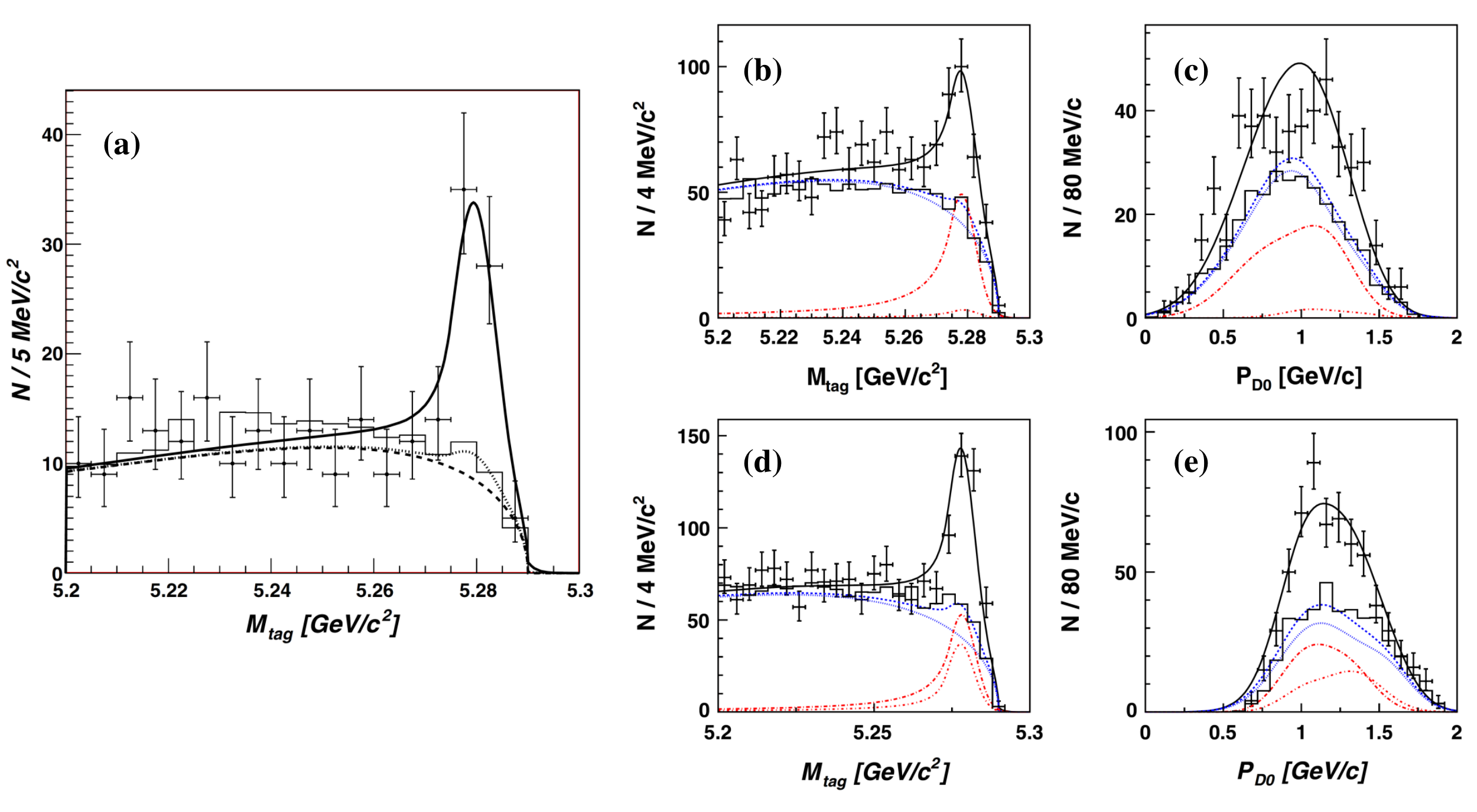}
}
\caption{The \mes\ (labeled $M_{\rm tag}$) distribution of
\BztoDsttaunu\ candidates at Belle{\protect\cite{Matyja:2007kt}} (a). 
The histogram shows the total expected background, and the dashed,
dotted, and solid curves are the contributions to the fit from the 
combinatorial background, total background, and background plus signal,
respectively.
The \mes\ (b) and $D^0$ momentum (c) distributions of \BptoDsttaunu\
candidates and (d,e) \BptoDtaunu\
candidates at Belle{\protect\cite{Bozek:2010xy}}.  The solid curve shows the
total fit function, and the other cuves show the fit contributions of
the combinatorial background, total background, and \BptoDsttaunu\ and
\BptoDtaunu\ signals. {\protect\footnotemark}
\protect\label{fig:bll-Dtaunu-results} }
\end{figure}

Peaking background, defined to be non-signal events with a peaking
\mes\ distribution and arising mostly from $B^0\to D^{*-}e^+\nu_e$,
was determined from simulation to constitute about 6 events.
The signal and combinatorial-background yields were determined with a
fit to the \mes\ distribution.
The signal yield was $60 ^{+12}_{-11}$ events,
with a significance\footnote{All quoted signal significances account 
  for the relevant systematic uncertainties.} 
of $5.2\sigma$. The branching fraction was measured to be
\beq
\br(\BztoDsttaunu) = \left(2.02 ^{+0.40}_{-0.37} \pm 0.37 \right)\%.
\label{eq:bll-2007-dtaunu}
\eeq
%
The \mes\ distribution and the overlaid fit function are shown in 
Fig.~\ref{fig:bll-Dtaunu-results}(a).

\footnotetext{Fig.~{\protect\ref{fig:bll-Dtaunu-results}}(a) is reprinted with
permission from A. Matyja {\it et al.}, Phys. Rev. Lett. {\bf 99}, 191807
(2007). Copyright (2007) by the American Physical Society.
Figs.~{\protect\ref{fig:bll-Dtaunu-results}}(b-e) are reprinted with permission
from A. Bozek {\it et al.},   Phys.\ Rev.\ Lett.\  {\bf 99}, 191807 (2007). 
Copyright (2010) by the American Physical Society.}

In 2010, Belle reported a study of \BptoDsttaunu\ and \BptoDtaunu\ with
the same analysis technique and a larger data sample of $657\times
10^6$ $B\bar B$ pairs\cite{Bozek:2010xy}.
In addition to the signal decay modes used for the 2007
analysis\cite{Matyja:2007kt}, the decays $\bar D^{*0}\to \bar
D^0\pi^0$, $\bar D^{*0}\to \bar D^0\gamma$, and $\tau^+\to
\mu^+\nu_\mu \bar\nu_\tau$ were used.  
Fits to the kinematic-variable distributions of 
a data sample selected with signal-rejection requirements 
were used to determine the relative
contributions of different background sources. 
The signal and combinatorial-background yields were obtained from
a fit to \mes\ and the $D^0$ momentum in the CM frame. 
The signal yields for \BptoDsttaunu\ and \BptoDtaunu\
were $446 ^{+58}_{-56}$ 
and $146 ^{+42}_{-41}$  events, respectively,
with branching fractions of 
\beqa
\br(\BptoDsttaunu) = \left(2.12 ^{+0.28}_{-0.27} \pm 0.29 \right)\%
     \nonumber\\
\br(\BptoDtaunu) = \left(0.77 \pm 0.22 \pm 0.12 \right)\%.
\label{eq:bll-2010-dtaunu}
\eeqa
The decay \BptoDsttaunu\ was observed with a significance of
$8.1\sigma$, and evidence for \BptoDtaunu\ was established at $3.5\sigma$.
The event distributions and fits are shown in
Fig.~\ref{fig:bll-Dtaunu-results}(b-e).

\subsubsection{Belle hadronic-tagging Measurement}
\label{sec:dtaunu-belle-hadtag}

Belle performed a hadronic-tagging analysis of the four
channels \BztoDtaunu\, \BztoDsttaunu, \BptoDtaunu, and \BptoDsttaunu,
using $657\times 10^6$ $B\bar B$ pairs.
The results were reported at a conference\cite{Adachi:2009qg} in 2009,
but have not been published. 
Signal-$B$ reconstruction was performed in the two leptonic decays $\tau^+\to
\ell^+\nu_\ell\bar\nu_\tau$, a total of 10 $D$-meson decay
modes, and 4 $D^*$ modes.
The tag-$B$ was reconstructed only in
the two-body decays $\bar B \to D^{(*)} h^-$, 
where $h^-$ was a $\pi^-$, $\rho^-$, $a_1^-$, or $D_s^{(*)-}$,
with a total of 15 $D_{(s)}$ decay modes and 5 $D_{(s)}^*$ modes.

\begin{figure}[!btph]
\centerline{
\includegraphics[width=1.1\textwidth]{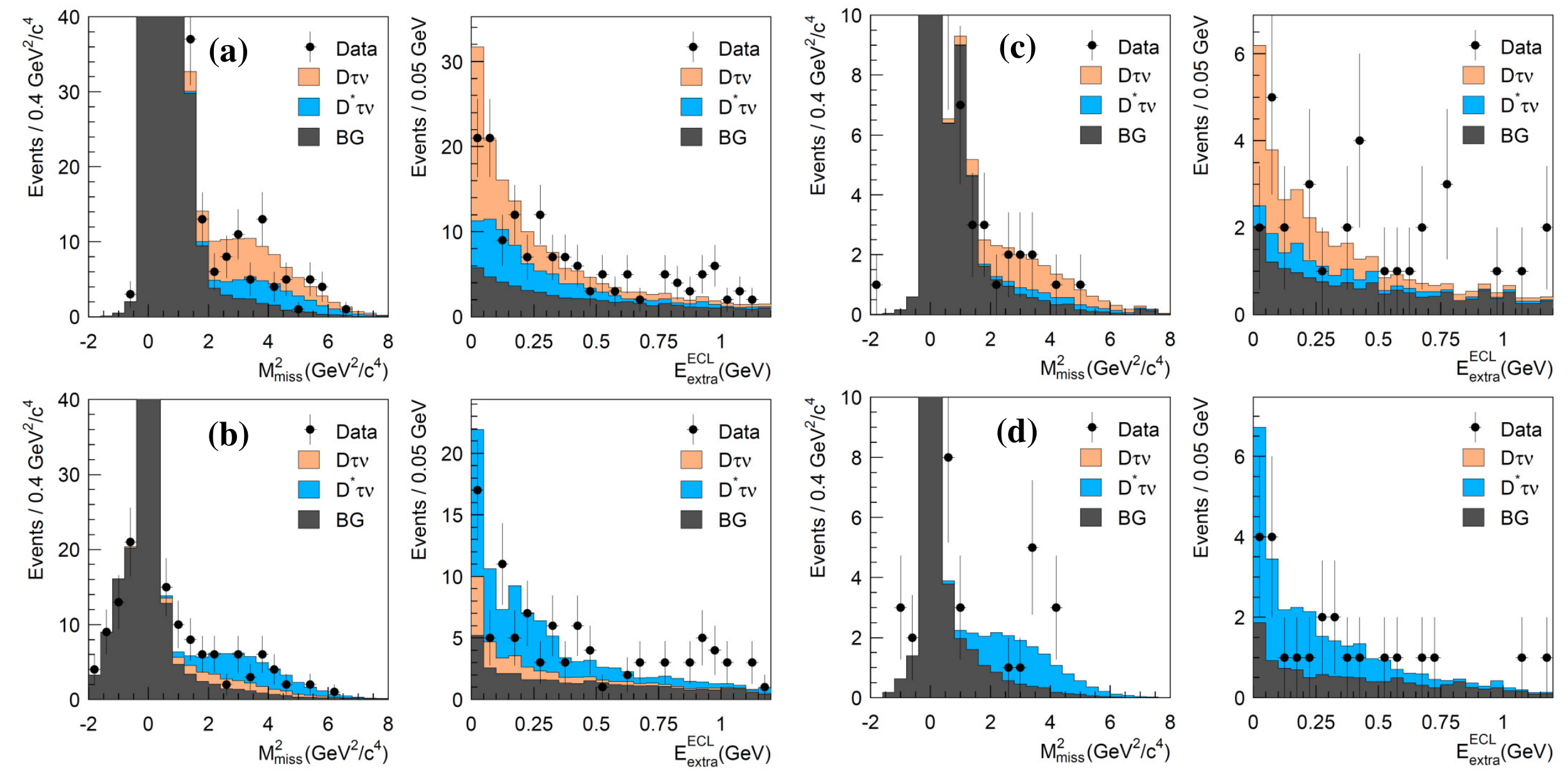}
}
\caption{Distributions of \mmisssq\ and \Eex\ 
(the left and right figures, respectively,
in each labeled pair of plots) for 
\BptoDtaunu\ (a), \BptoDsttaunu\ (b), 
\BztoDtaunu\ (c), and \BztoDsttaunu\ (d)
candidates in the preliminary Belle
analysis{\protect\cite{Adachi:2009qg}}.
Shaded histograms show the fit results.
\protect\label{bll-Dtaunu-results-conf}
}
\end{figure}

The yield of \BtoDstlnuGen\ events in each of the four modes was
obtained by fitting the \mmisssq\ distribution in the control sample
defined by $|\mmisssq|<1~\gev^2$. The signal yields were obtained from
a two-dimensional fit to the distribution of \mmisssq\ vs. \Eex\ in
the range $-2 < \mmisssq < 8~\gev^2$. The two fit variables were found
to be uncorrelated for \BtoDsttaunuGen\ and \BtoDstlnuGen, and the
correlation for the remaining background types was accounted for using
simulated events. The distributions of these variables and the fit
functions are shown in Fig.~\ref{bll-Dtaunu-results-conf}.
The ratio $\RDstGen$ was extracted from the two yields, accounting for
the different \BtoDstlnuGen\ efficiencies in the two samples.  The
results of the fits are summarized in
Table~\ref{tab:bll-dtaunu-results-2009}.

%
\begin{table}[htbp]
\tbl{Results of the preliminary \BtoDsttaunuGen\ analysis from
Belle{\protect\cite{Adachi:2009qg}}, showing for each mode the number
of signal events, the ratio $\RDstGen$, the branching fraction, and
the signal significance. 
Where given, the third uncertainty is due to the branching fraction 
$\br(\BtoDstlnuGen)$.}
{\begin{tabular}{lcccr} 
\toprule 
Decay mode & $N_{\rm signal}$ & $\RDstGen$ & $\br (\%)$ & Significance \\
\colrule 
\BptoDtaunu   & $98.6 ^{+26.3}_{-25.0}$ & $0.70 {^{+0.19}_{-0.18}} {^{+0.11}_{-0.09}}$ 
              & $1.51 {^{+0.41}_{-0.39}} {^{+0.24}_{-0.19}} \pm 0.15$ & 3.8 \\[3pt]
\BztoDtaunu   & $17.2 ^{+7.7}_{-6.9}$   & $0.48 {^{+0.22}_{-0.19}} {^{+0.06}_{-0.05}}$ 
              & $1.01 {^{+0.46}_{-0.41}} {^{+0.13}_{-0.11}} \pm 0.10$ & 2.6 \\[3pt]
\BptoDsttaunu & $99.8 ^{+22.2}_{-21.3}$ & $0.47 {^{+0.11}_{-0.10}} {^{+0.06}_{-0.07}}$ 
              & $3.04 {^{+0.69}_{-0.66}} {^{+0.40}_{-0.47}} \pm 0.22$ & 3.9 \\[3pt]
\BztoDsttaunu & $25.0 ^{+7.2}_{-6.3}$   & $0.48 {^{+0.14}_{-0.12}} {^{+0.06}_{-0.04}}$ 
              & $2.56 {^{+0.75}_{-0.66}} {^{+0.31}_{-0.22}} \pm 0.10$ & 4.7 \\
\botrule
\end{tabular}{\protect\label{tab:bll-dtaunu-results-2009}}}
\end{table}

\subsubsection{\babar\ Hadronic-Tagging Measurement}
\label{sec:dtaunu-babar-hadtag}

In 2008,  \babar\  reported the first study of the four
\BtoDsttaunuGen\ channels 
using a sample of $232\times 10^6$ $B \bar
B$ pairs.  The decay \BptoDsttaunu\ was observed with a significance
of $5.3\sigma$, and evidence for \BztoDtaunu\ was obtained at
$3.3\sigma$.  Rather than describing this analysis in detail, we do so
for the 2012 analysis\cite{ref:bbr-dtaunu-2012,ref:bbr-dtaunu-2013}
that superseded it, and which used a larger data
sample ($471\times 10^6$ $B\bar B$ pairs) and improved hadronic tagging.
(see discussion in Sec.~\ref{sec:hadronic-tag}).

Tag-$B$ reconstruction in the 2012 analysis was performed with
1680 final states.
Signal-$B$ decays were reconstructed in the two $\tau^+$ 
leptonic modes, 11 $D$ modes, and four $D^*$ modes.
The final analysis stage was a simultaneous fit to the two-dimensional
distributions of $\mmisssq$ vs. the lepton momentum $p_\ell^*$ in the
signal-$B$ rest frame. The correlation between these variables
necessitated evaluation of the two-dimensional fit functions
from simulated events.
The fit was performed simultaneously on the four
\BtoDsttaunuGen\ candidate samples plus four control samples, in which an
additional $\pi^0$ was reconstructed in an attempt to identify
production of a $D^{**}$, defined here as an excited charm state heavier
than the $D^*$. The control samples helped
determine the contribution of poorly understood $B\to \bar D^{**}
\ell^+ \nu_\ell$ and $B\to \bar D^{**} \tau^+ \nu_\tau$ backgrounds to
the \BtoDsttaunuGen\ candidate samples.
%
%
The ratios of the \BtoDsttaunuGen\ and \BtoDstlnuGen\ yields were used
to calculate $\RDstGen$, and earlier \babar\
meausrements\cite{Aubert:2009ac,Aubert:2008yv,Aubert:2007rs} of
$\br(\BtoDstlnuGen)$ were used to obtain $\br(\BtoDsttaunuGen)$.

\begin{figure}[!btph]
\centerline{
\includegraphics[width=1.1\textwidth]{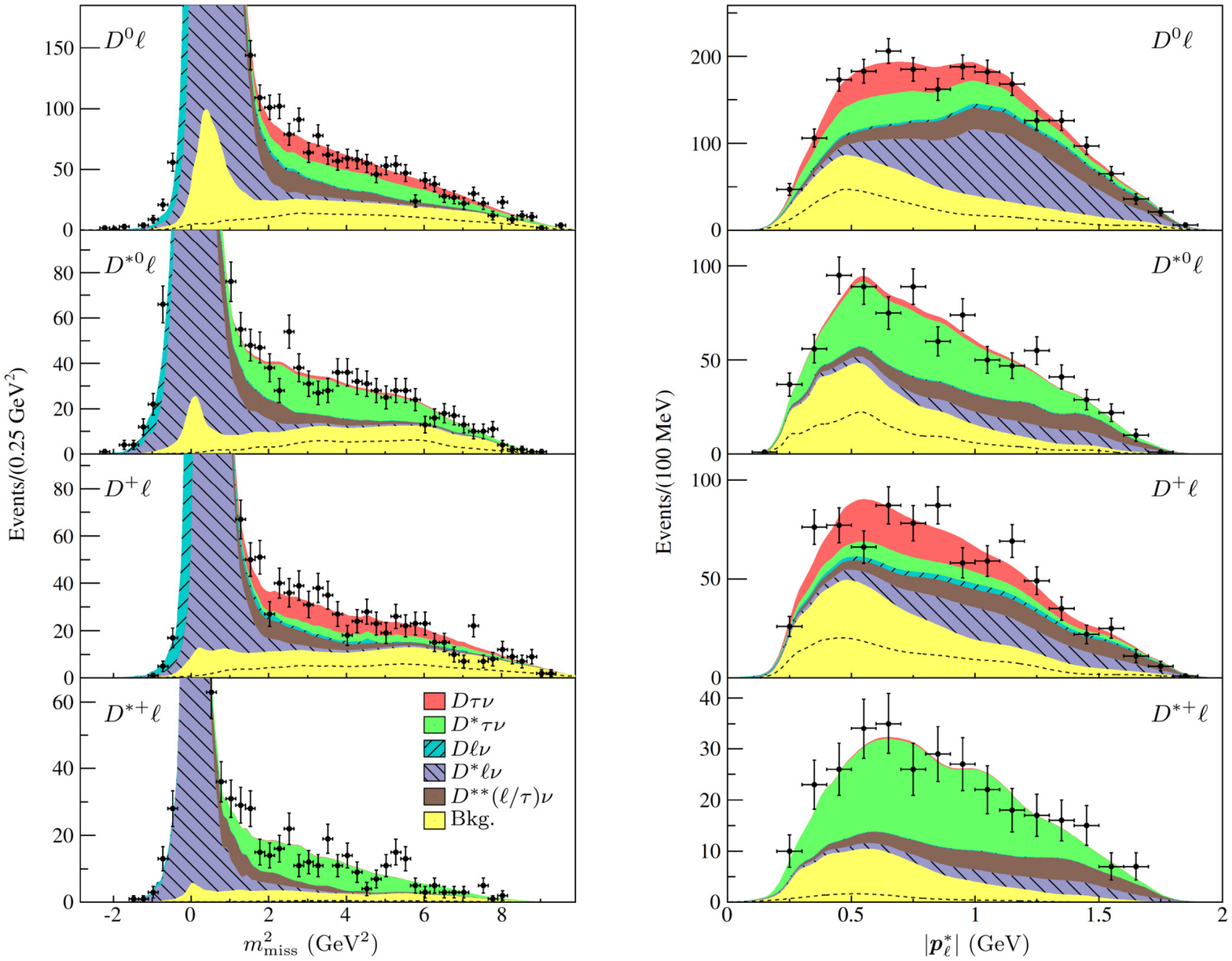}
}
\caption{\mmisssq\ (left) and $p_l^*$ (right) distributions of the
\BtoDsttaunuGen\ candidates reconstructed by
\babar{\protect\cite{ref:bbr-dtaunu-2012,ref:bbr-dtaunu-2013}}. 
Shaded regions show the
results of the fit with the isospin constraint $R(D^{(*)0}) =
R(D^{(*)+}) \equiv \RDstGen$. The reconstructed final state is shown on
each plot. 
The $p_l^*$ distributions were produced with the requirement $\mmisssq
>1$~GeV to suppress the large \BtoDstlnuGen\ peak, which is truncated
in the \mmisssq\ distributions.  The dashed line shows the level of
the continuum background.  \protect\label{bbr-Dtaunu-results} }
\end{figure}

The analysis resulted in the first significant observation of
\BtoDtaunu. 
The \mmisssq\ and $p_l^*$ distributions of the data are shown in
Fig.~\ref{bbr-Dtaunu-results}, overlaid with the fit function
when the isospin constraint $R(D^{(*)0}) = R(D^{(*)+}) \equiv \RDstGen$
was applied. 
The results are presented in
Table~\ref{tab:bbr-dtaunu-results} for each of the four decay modes
and for the isospin-constrained fit.

\babar\ found the measured values of \RD\ and \RDst\ to be higher by
$2.0\sigma$ and $2.7\sigma$, respectively, than the SM expectation
(Eq.~(\ref{eq:R-SM})). Accounting for correlations between the \RD\ and \RDst\
measurements, the combined consistency with the SM was $3.4\sigma$,
corresponding to a $p$-value of $6.9\times 10^{-4}$.  This is reduced
to $3.2\sigma$ when using Eq.~(\ref{eq:R-SM-lattice}),
with similar results obtained for Eq.~(\ref{eq:R-SM-mintheory}).
The measured branching fractions for the four modes
are higher than the predictions of Eq.~(\ref{eq:Dtaunu-brs}).  
by $1.4$, $1.7$, $1.3$, and $1.9$ standard deviations, respectively.
The $q^2$ spectra were found to be consistent with the SM
to within the statistical uncertainties.

%
\begin{table}[htbp]
\tbl{Results of the \BtoDsttaunuGen\ analysis from
\babartbl{\protect\cite{ref:bbr-dtaunu-2012,ref:bbr-dtaunu-2013}},
showing for each mode the number of signal events, the ratio
$\RDstGen$, the branching fraction, and the signal significance.}
{\begin{tabular}{lcccr} 
\toprule 
Decay mode & $N_{\rm signal}$ & $\RDstGen$ & $\br (\%)$ & Significance ($\sigma$)\\
\colrule 
\BptoDtaunu   & $314 \pm 60$ & $0.429 \pm 0.082 \pm 0052$ & $0.99 \pm 0.19 \pm 0.13$ & 4.7 \\ 
\BztoDtaunu   & $177 \pm 31$ & $0.469 \pm 0.084 \pm 0053$ & $1.01 \pm 0.18 \pm 0.12$ & 5.2 \\ 
\BptoDsttaunu & $639 \pm 62$ & $0.322 \pm 0.032 \pm 0022$ & $1.71 \pm 0.17 \pm 0.13$ & 9.4 \\ 
\BztoDsttaunu & $245 \pm 27$ & $0.355 \pm 0.039 \pm 0021$ & $1.74 \pm 0.19 \pm 0.12$ & 10.4 \\ 
\colrule
\BtoDtaunu    & $489 \pm 63$ & $0.440 \pm 0.058 \pm 0042$ & $1.02 \pm 0.13 \pm 0.11$ & 6.8 \\ 
\BtoDsttaunu  & $888 \pm 63$ & $0.332 \pm 0.024 \pm 0018$ & $1.76 \pm 0.13 \pm 0.12$ & 13.2 \\ 
\botrule
\end{tabular}{\protect\label{tab:bbr-dtaunu-results}}}
\end{table}

\subsubsection{\boldmath Summary and Consistency of \BtoDsttaunuGen\ Measurements}
\label{sec:dtaunu-results-summary}

As shown discussed above, the \BtoDsttaunuGen\ rate measurements have
consistently yielded results higher than the SM expectations.  Comparison
of theory and experimental results from both \babar\ and Belle is best
performed in terms of the branching-fraction ratios \RDstGen. However,
this is complicated by the fact that the published Belle results were
given in terms of the branching fractions, and correlations between
the \RD\ and \RDst\ results in the Belle measurements have not been
published. An unofficial combination of the
published\cite{Matyja:2007kt,Bozek:2010xy} and
preliminary\cite{Adachi:2009qg} Belle results with the
\babar\ results\cite{ref:bbr-dtaunu-2012,ref:bbr-dtaunu-2013} has been
performed\cite{ref:bozek-fpcp} in terms of \RDstGen. This calculation
found the combined Belle results for \RDstGen\ to be within
$3.3\sigma$ of the SM prediction. Combining results from both
experiments yielded a discrepancy of $4.8\sigma$ with respect to the
SM.

Lastly, we check the consistency of the $B$-factory results
with the ALEPH measurement\cite{Barate:2000rc} of the inclusive
branching fraction
\beq
\br(b_Z\to D^{*-} \tau^+\nu_\tau X) = (0.88 \pm 0.31 \pm 0.28)\%, 
\label{eq:aleph}
\eeq
where $X$ stands for possible additional particles, and $b_Z$
indicates a $b$ quark produced in $Z^0\to b\bar b$.  The fraction of
these quarks that hadronize into $B^+$ or $B^0$ mesons is $f_{Z\to B}
= (80.8 \pm 1.8)\%$, where equal production of both meson types is
assumed\cite{hfag-fracz}. The remaining $\sim 20\%$ hadronize into
$B_s$ mesons and $b$~baryons, which undergo semileptonic decays that
tend to produce $D_s$ mesons and charmed baryons,
respectively\cite{pdg12}, rather than $D^{*-}$ mesons.  Therefore, the
dominant source of $D^{*-} \tau^+\nu_\tau X$ events in the ALEPH
measurement was $B^+$ and $B^0$ decays. We note that\cite{pdg12}
\beq
\br(B^0\to D^{*-}\ell^+\nu_\ell) \approx \br(B^+\to
D^{(*)}(n\pi)\ell^+\nu_\ell) + \br(B^0\to
D^{(*)}(n\pi)\ell^+\nu_\ell), 
\label{eq:brs-npi}
\eeq
where $(n\pi)$ stands for at least one pion. Assuming this approximate
relation holds for decays with a $\tau^+$ lepton in the final state,
one obtains the expectation
\beq
\br(b_Z\to D^{*-} \tau^+\nu\tau X) \approx 
{f_{Z\to B} \over 2} \left(1 + F_{D^{*-}} {R(D^* (n\pi)) \over \RDst}\right)
\br(\BztoDsttaunu),
\label{eq:btoDtaunu}
\eeq
where $F_{D^{*-}}$ is the fraction of decays on the right-hand side of
Eq.~(\ref{eq:brs-npi}) in which a $D^{*-}$ is produced, and $R(D^*
(n\pi)) \equiv \br(B \to D^{(*)}(n\pi)\tau^+\nu_\tau) / \br(B \to
D^{(*)}(n\pi)\ell^+\nu_\ell)$.  Given the fraction of $D^{*-}$
production in $B\to \bar D^{(*)}\ell^+ \nu_\ell$ decays\cite{pdg12},
we take $F_{D^{*-}}$ to be between $1/4$ and $1/2$. 
Phase-space considerations suggest $R(D^* (n\pi)) < \RDst$,
but to be conservative, we take this relation to be an equality.
Then with the value of $\br(\BztoDsttaunu)$
from Table~\ref{tab:bbr-dtaunu-results}, Eq.(\ref{eq:btoDtaunu}) predicts
$\br(b_Z\to D^{*-} \tau^+\nu\tau X)$ to be between $0.9$ and $1.0$, with
the range being due to our choices for $F_{D^{*-}}$. This is in excellent
agreement with the measured value, Eq.(\ref{eq:aleph}).

\subsection{\Btotaunu\ Measurements}
\label{sec:taunu}

Prior to the start of the $B$~factory programs, searches for 
\Btotaunu  were conducted by ARGUS\cite{Albrecht:1994jg},
CLEO\cite{Artuso:1995ar,Browder:2000qr},
ALEPH\cite{Buskulic:1994gj,Barate:2000rc}, and
L3\cite{Acciarri:1996bv}, reaching a limit of 
$\br(\Btotaunu) < 5.7\times 10^{-4}$.
Between 2004 and 2013, \babar\ and Belle published a total of nine
papers on the topic, using both semileptonic
tagging\cite{Aubert:2004kz,Aubert:2005du,Aubert:2007bx,Aubert:2009wt,Hara:2010dk}
and hadronic
tagging\cite{Ikado:2006un,Aubert:2007xj,ref:bll-taunu-2013,Lees:2012ju}.
First evidence for this decay, at a level of $3.5\sigma$, was obtained
by Belle with hadronic tagging\cite{Ikado:2006un} and a data sample
containing $449\times 10^6$ $B\bar B$ pairs, and resulted in a
branching-fraction measurement of $\br(\Btotaunu) = (1.79
{^{+0.56}_{-0.49}} {^{+0.45}_{-0.51}})\times 10^{-4}$.  Results became
more precise as the data samples grew and analysis methods improved.

In what follows, we describe the four most recent $B$-factory
measurements of $\br(\Btotaunu)$. A summary of the experimental
results and how they compare to the SM expectation is given in
Sec.~\ref{sec:taunu-summary}.

\subsubsection{Semileptonic-Tagging Measurements}

\begin{figure}[htbp]
\addtocounter{footnote}{-1}
\centerline{
\includegraphics[width=1.0\textwidth]{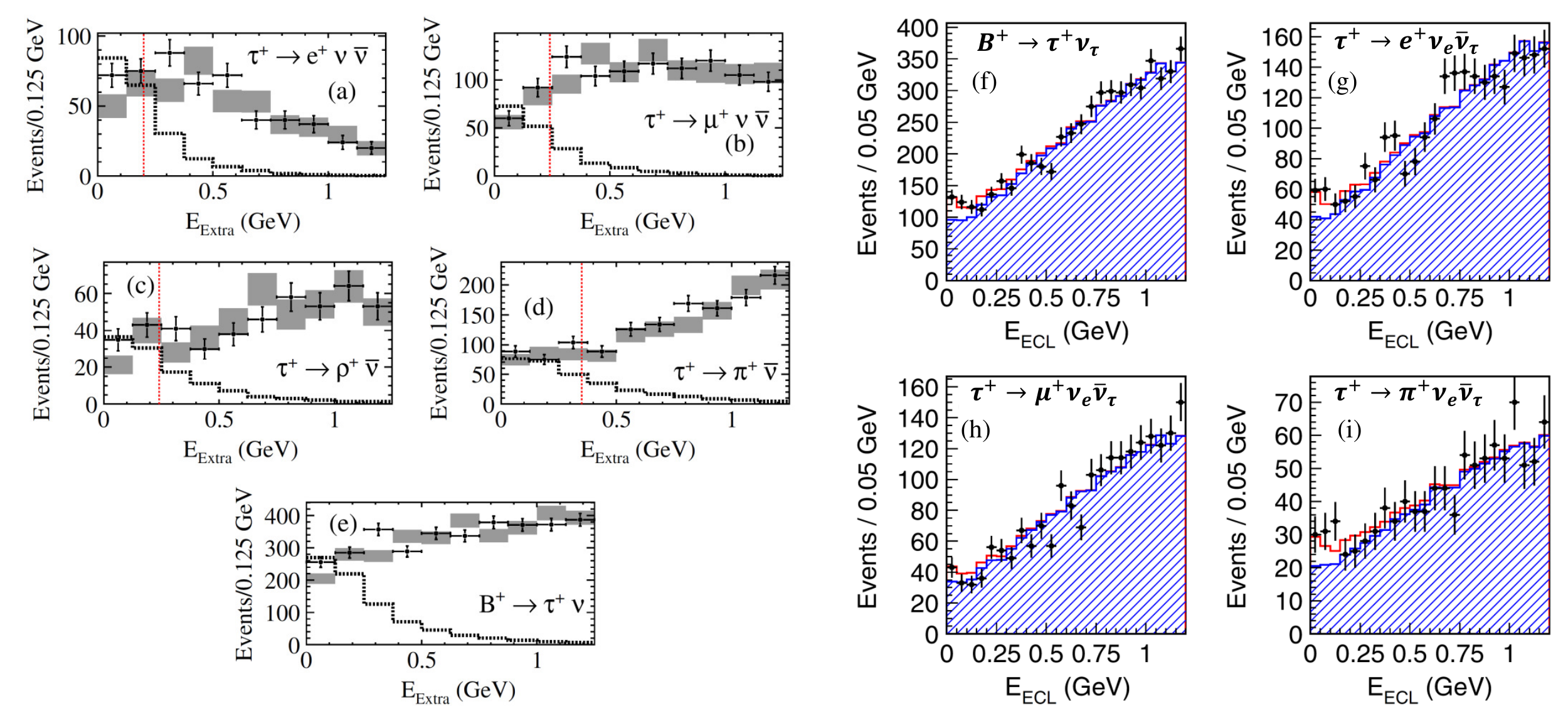}
}
\caption{\Eex\ distributions in the \babar\ semileptonic-tagging
\Btotaunu\ analysis{\protect\cite{Aubert:2009wt}}, shown for each 
$\tau^+$ decay mode (a-d) and for the sum of the modes (e).  The
grey boxes show the background expectation from simulation, normalized
to the sideband $\Eex < 0.4~\gev$ indicated by the dotted vertical
line. The dotted histogram is ten times the expected signal contribution.
Also shown are \Eex\ distributions in the corresponding Belle
analysis{\protect\cite{Hara:2010dk}} for each mode (g-i) and 
for the sum of the modes (f). The hatched blue histogram is the background
expectation,≈2 and the solid red histogram includes the signal contribution.
{\protect\label{fig:taunu-SLtag-2010}}}
\end{figure}

In 2010, \babar\ and Belle published studies of \Btotaunu\ with
semileptonic tagging. 
The \babar\ analysis\cite{Aubert:2009wt} used a data sample of
$459\times 10^6$ $B \bar B$ pairs.  They reconstructed the tag~$B$ in
the decays $B^-\to D^0 \ell^-\bar\nu_\ell X$, where $X$ stands for
possible additional particles that were not reconstructed.
The $\tau^+$ was reconstructed in the leptonic decays 
$\tau^+\to \ell^+ \nu_\ell \bar\nu_\tau$ and the hadronic decays
$\tau^+\to \pi^+\bar\nu_\tau$ and $\tau^+\to \rho^+\bar\nu_\tau$.
The signal yield in each $\tau^+$ channel was measured from the number
of events in the signal region $\Eex < 0.4~\gev$, after subtraction of
the expected background yield. This, in turn, was obtained from the simulated
\Eex\ distribution, normalized to the sideband $\Eex > 0.4~\gev$.
%
%
The simulation predictions for the \Eex\ distributions of the
background were validated using a double-tag control sample, in which both 
$B$ mesons were reconstructed via semileptonic decays.
\babar\ observed 583 signal-region events with a background expectation of 
$509 \pm 30$ events, and reported the branching fraction 
$\br(\Btotaunu) = (1.7 \pm 0.8 \pm 0.2)\times 10^{-4}$,
with a signal significance of $2.3\sigma$.
A mode-by-mode breakdown of the results is shown in 
Table~\ref{tab:taunu-SLtag}, and 
the \Eex\ distributions are shown in Figs.~\ref{fig:taunu-SLtag-2010}(a-e).

\begin{table}[htbp]
\tbl{Results of the semileptonic-tagging \Btotaunu\ analyses from
\babartbl{\protect\cite{Aubert:2009wt}} and
Belle{\protect\cite{Hara:2010dk}}, showing 
the expected number of
background events ($N_{\rm background}$) and the number of observed events
($N_{\rm observed}$) in the signal region for \babartbl, the number of
signal events ($N_{\rm signal}$) obtained from the fit for Belle, and
the branching fraction $\br(\Btotaunu)$ for both experiments.}
{\begin{tabular}{l|ccc|cc}
\hline
&&&&&\\[-6pt]
Decay mode & \multicolumn{3}{c}{\babar\ results}& \multicolumn{2}{|c}{Belle results} \\

& $N_{\rm background}$ & $N_{\rm observed}$ & $\br (\times 10^{-4})$& $N_{\rm signal}$ & $\br (\times 10^{-4})$ \\[3pt]
\hline
&&&&&\\[-6pt]
$\tau^+\to e^+\nu_e\bar\nu_\tau$    & $81 \pm 12$ & $121$ & $3.6\pm 1.4$ & $73 ^{+23}_{-22}$ & $1.90 {^{+0.59}_{-0.57}} {^{+0.33}_{-0.35}}$ \\[3pt]

$\tau^+\to \mu^+\nu_e\bar\nu_\tau$  & $135 \pm 13$ & $148$ & $1.3 ^{+1.8}_{-1.6}$ & $12 ^{+18}_{-17}$ & $0.5 {^{+0.76}_{-0.72}} {^{+0.18}_{-0.21}}$ \\[3pt]

$\tau^+\to \pi^+\nu_e\bar\nu_\tau$  & $234 \pm 19$ & $243$ & $0.6 ^{+1.4}_{-1.2}$ & $55 ^{+21}_{-20}$ & $1.80 {^{+0.69}_{-0.66}} {^{+0.36}_{-0.37}}$ \\[3pt]

$\tau^+\to \rho^+\nu_e\bar\nu_\tau$ & $59 \pm 9$ & $71$ & $2.1 ^{+2.0}_{-1.8}$ & & \\[3pt]
\hline
&&&&&\\[-6pt]

Combined & $509 \pm 30$ & $583$ & $1.7 \pm 0.8 \pm 0.2$ & $143 ^{+36}_{-35}$ & $1.54 {^{+0.38}_{-0.37}} {^{+0.29}_{-0.31}}$ \\[3pt]
\hline
\end{tabular}{\protect\label{tab:taunu-SLtag}}}
\end{table}

%
\footnotetext{Figs.~{\protect\ref{fig:bll-Dtaunu-results}}(f-i) are
reprinted with permission from K.~Hara {\it et al.}, Phys.\ Rev.\ D
{\bf 82}, 071101 (2010). Copyright (2010) by the American Physical
Society.}

The Belle semileptonic-tagging analysis used a data sample of
$657\times 10^6$ $B\bar B$ pairs. 
The $\tau^+$ was reconstructed in $\tau^+\to
\ell^+\nu_\ell \bar\nu\tau$ and $\tau^+\to \pi^+\bar\nu\tau$. 
A fit to the \Eex\ distribution provided the
yields of signal and background events for each $\tau^+$ decay
mode. The fit functions were obtained from simulation, corrected 
using control samples of double-tag events and data taken off the \fours\
resonance.
The \Eex\ distributions and fit functions are shown in
Fig.~\ref{fig:taunu-SLtag-2010}(f-i), and the signal yield and
branching fraction obtained for each mode are listed in
Table~\ref{tab:taunu-SLtag}.  Combining the four $\tau^+$ mode, Belle
found $143 ^{+36}_{-35}$ signal events, 
a signal significance of $3.6\sigma$,
and a branching-fraction measurement of 
$\br(\Btotaunu) = (1.54 \ ^{+0.48}_{-0.37} \ ^{+0.29}_{-0.31})\times 10^{-4}$.

\subsubsection{Hadronic-Tagging Measurements}
\label{sec:taunu-hadtag}

In 2013, \babar\ and Belle published \Btotaunu\ results based on their
full data sets and improved hadronic tagging methods (see
Sec.~\ref{sec:hadronic-tag}), leading to significant improvements over
previous hadronic-tagging results. Both analyses used the four decay
modes $\tau^+ \to \ell^+ \nu_\ell \bar\nu_\tau$, $\pi^+\bar\nu_\tau$,
and $\rho^+\bar\nu_\tau$.

\begin{figure}[htbp]
\addtocounter{footnote}{-1}
\centerline{
\includegraphics[width=1.0\textwidth]{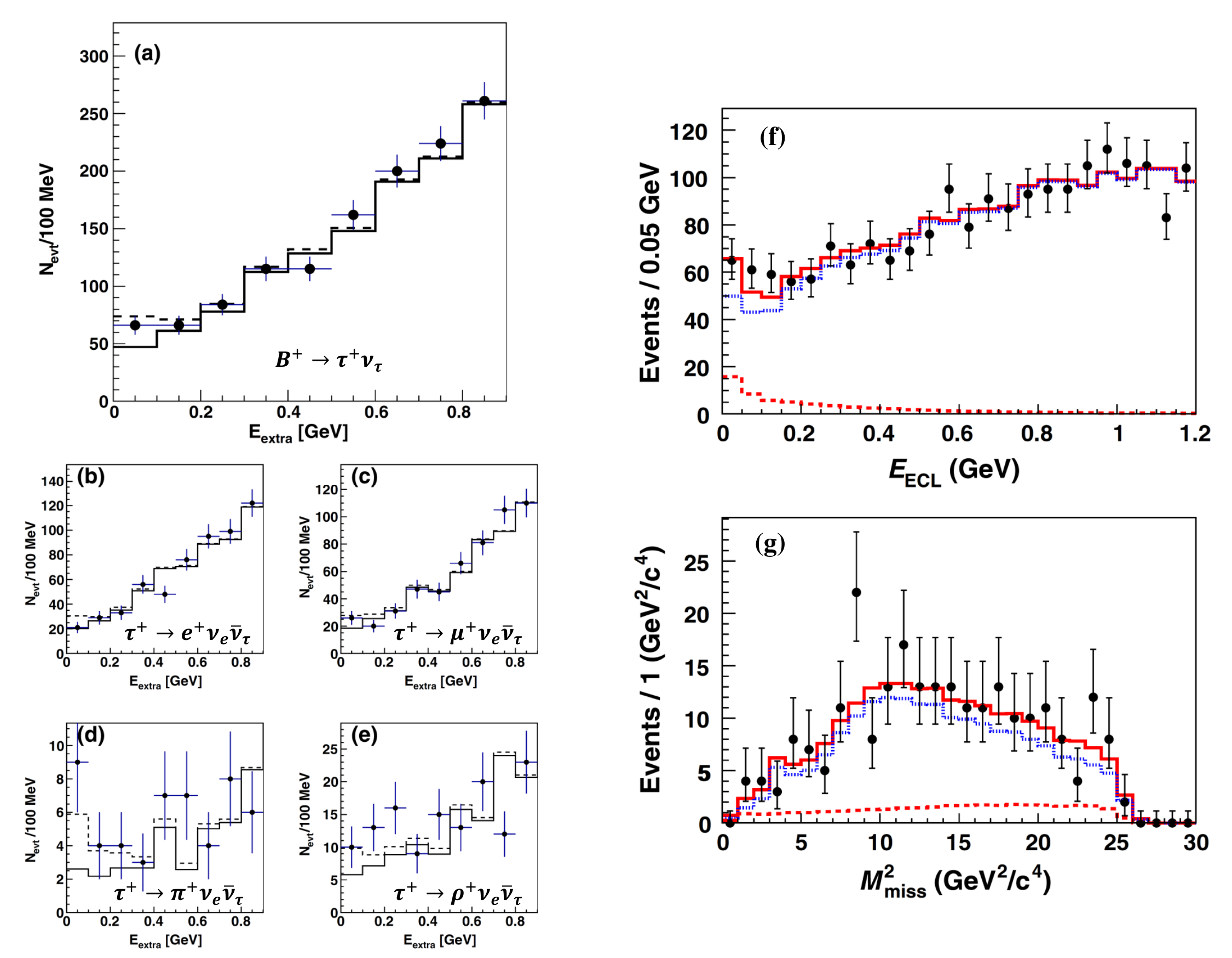}
}
\caption{\Eex\ distributions in the \babar\ hadronic-tagging
\Btotaunu\ analysis{\protect\cite{Lees:2012ju}}, shown for the sum of
the $\tau^+$ modes (a) and for each mode separately (b-e).  The 
solid histograms show the background contributions, and the dashed
histograms show the background-plus-signal fit functions.
The \Eex\ (f) and \mmisssq\ (g) distributions in the corresponding Belle
analysis{\protect\cite{ref:bll-taunu-2013}} are shown for the sum of the modes.
The dotted blue histograms are the background contribution, the 
red dashed histograms show the signal contribution, and the 
solid red is the total fit function.
{\protect\label{fig:taunu-hadtag-2013}}}
\end{figure}

The \babar\ analysis\cite{Lees:2012ju} was performed with a data
sample of $467.8\times 10^6$ $B \bar B$ pairs. 
A simultaneous fit to the \Eex\ distributions of all modes
was used to extract the signal branching fraction and the background
yield in each mode. 
The fit functions for events with a correctly reconstructed \btag\
were taken from simulation after corrections for data-simulation
discrepancies obtained from double-tag events, in which the signal~$B$
was replaced by a $B$~meson reconstructed via a hadronic or
semileptonic decay.  The fit functions for the combinatorial background
were histograms of data events in the sideband $5.209 < \mes <
5.260~\gev$.
The \Eex\ distributions are shown in Fig.~\ref{fig:taunu-hadtag-2013},
and the results are summarized in Table~\ref{tab:taunu-hadtag}.
\babar\ found a total of $62.1 \pm 17.3$ signal events and a
significance of $3.8\sigma$, and measured the branching fraction
$\br(\Btotaunu) = (1.83 ^{+0.53}_{-0.49} \pm 0.24)\times 10^{-4}$.

\begin{table}[htbp]
\tbl{Results of the hadronic-tagging \Btotaunu\ analyses from
\babartbl{\protect\cite{Lees:2012ju}}
and Belle{\protect\cite{ref:bll-taunu-2013}}, showing the signal
yield and the calculated branching fraction for each $\tau^+$
mode and for the combination of the modes.}
{\begin{tabular}{l|cc|cc} 
\hline
Decay mode & \multicolumn{2}{c}{\babartbl\ results} 
               & \multicolumn{2}{|c}{Belle results} \\
           & $N_{\rm signal}$ & $\br (\times 10^{-4})$ 
               & $N_{\rm signal}$ & $\br (\times 10^{-4})$ \\
\hline
&&&&\\[-6pt]
$\tau^+\to e^+\nu_e\bar\nu_\tau$    & $4.1 \pm 9.1$ & $0.35 ^{+0.84}_{-0.73}$ 
               & $16 ^{+11}_{-9}$ & $0.68 ^{+0.49}_{-0.41}$ \\[3pt]
$\tau^+\to \mu^+\nu_e\bar\nu_\tau$  & $12.9 \pm 9.7$ & $1.12 ^{+0.90}_{-0.78}$ 
               & $26 ^{+15}_{-14}$ & $1.06 ^{+0.63}_{-0.58}$ \\[3pt]
$\tau^+\to \pi^+\nu_e\bar\nu_\tau$  & $17.1 \pm 6.2$ & $3.69 ^{+1.42}_{-1.22}$ 
               & $8 ^{+10}_{-8}$ & $0.57 ^{+0.70}_{-0.59}$ \\[3pt]
$\tau^+\to \rho^+\nu_e\bar\nu_\tau$ & $24.0 \pm 10.0$ & $3.78 ^{+1.65}_{-1.45}$ 
               & $14 ^{+19}_{-16}$ & $0.52 ^{+0.72}_{-0.62}$ \\[3pt]
\hline
&&&&\\[-6pt]
Combined & $62.1 \pm 17.3$ & $1.83 ^{+0.53}_{-0.49} \pm 0.24$ 
               & $62 ^{+23}_{-22}$ & $0.72 ^{+0.27}_{-0.25} \pm 0.11$ \\[3pt]
\hline
\end{tabular}{\protect\label{tab:taunu-hadtag}}}
\end{table}

%
\footnotetext{Figs.~{\protect\ref{fig:bll-Dtaunu-results}}(f-g) are
reprinted with permission from I.~Adachi {\it et al.}, Phys.\ Rev.\
Lett.\ {\bf 110}, 131801 (2013). Copyright (2013) by the American
Physical Society.}

The Belle hadronic-tagging analysis\cite{ref:bll-taunu-2013} made use of
$772\times 10^6$ $B\bar B$ pairs. 
The signal yield was obtained from a two-dimensional fit to the
distribution of \Eex\ vs. \mmisssq. The distributions of the two
variables were found to be uncorrelated, except for $\tau^+\to
\rho^+\bar\nu_\tau$ events reconstructed as $\tau^+\to
\pi^+\bar\nu_\tau$, for which the correlation was taken into account
in the fit function. Double-tagged events were used to validate the 
signal fit functions.
The \Eex\ and \mmisssq\ distributions of the data and the corresponding
fit functions are shown in Fig.~\ref{fig:taunu-hadtag-2013}, and the 
fit results are shown in Table~\ref{tab:taunu-hadtag}.
The total signal yield was $62 ^{+23}_{-22}$  events, and the branching
fraction was found to be $\br(\Btotaunu) = 0.72 ^{+0.27}_{-0.25} \pm 0.11$,
with a signal significance of $3.0\sigma$.

\subsubsection{\boldmath Summary of \Btotaunu\ Results}
\label{sec:taunu-summary}
 
As the $B$-factory data samples grew and $\br(\Btotaunu)$ 
results became more precise,
tension was building between the experimental average and the SM
expectations based on $|V_{ub}|$ from exclusive semileptonic decays or on the
unitarity-triangle fits. For example, the
CKMfitter\cite{ref:ckmfitter} expectation for $\br(\Btotaunu)$,
Eq.~(\ref{eq:taunu-SM-fits}), differed by $2.6\sigma$ from the
experimental world average of $(1.65 \pm 0.34)\times 10^{-4}$ before
the 2013 hadronic-tagging measurement from
Belle\cite{ref:bll-taunu-2013}. This new measurement now dominates the
world average of $(1.15 \pm 0.23)\times 10^{-4}$, which is only
$1.7\sigma$ from the CKMfitter expected value.
While the new world average is $2.4\sigma$ from the predicted value of
$R'$ (Eq.~(\ref{eq:R'})), at the more relevant high values of $q^2$,
the difference is reduced\cite{Fajfer:2012jt} to $1.6\sigma$.

\begin{figure}[htbp]
\centerline{
\includegraphics[width=1.0\textwidth]{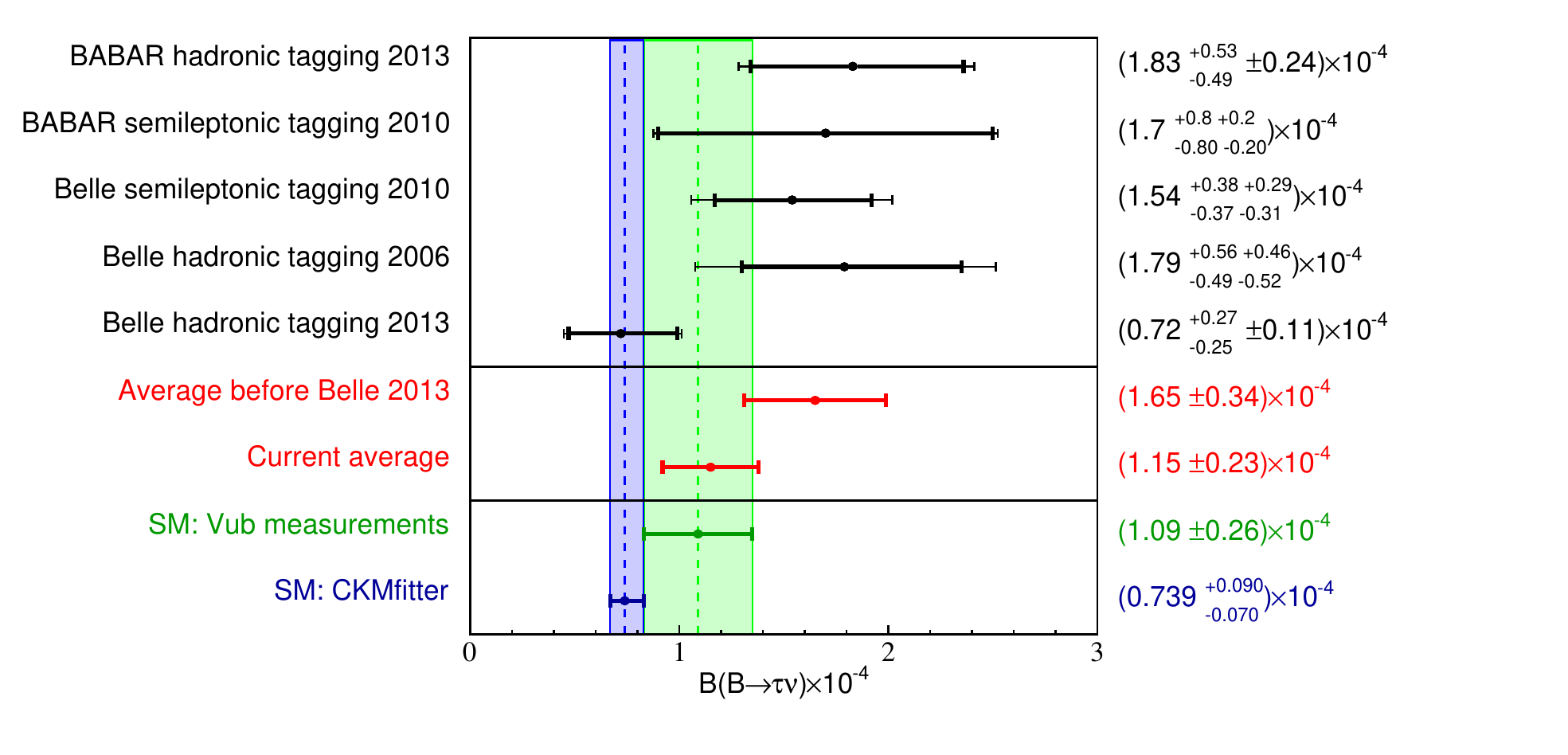}
}
\caption{Summary of the latest $\br(\Btotaunu)$ measurements, with the
  experiment, tagging method, and
  publication year indicated.  Inner (outer) error bars show the
  statistical (total) uncertainties.  Also shown are the PDG world
  average{\protect\cite{pdg12}} before the 2013 Belle
  measurement{\protect\cite{ref:bll-taunu-2013}} and the current PDG
  average, which was substantially lowered by that measurement. Two SM
  expectations are shown as vertical bands: using either the PDG
  average of $|V_{ub}|$ measurements
  (Eq.~(\protect\ref{eq:taunu-SM-PDG})) or the CKMfitter value
  (Eq.~(\protect\ref{eq:taunu-SM-fits}))
  {\protect\label{fig:BtotaunuSummary}}. }
\end{figure}

Fig.~\ref{fig:BtotaunuSummary} summarizes the results and their
averages before and after the 2013 Belle measurement, comparing them them with 
two SM expectation values: one based on $V_{ub}$ from the direct
measurement, and the other from CKMfitter.
Finally, we note that the na\"{\i}ve significance of the world
average for $\br(\Btotaunu)$ is $1.15/0.23 = 5\sigma$.

%

\section{New-Physics Interpretation of the Results} 
\label{sec:interp}

Given that $\br(\Btotaunu)$ has now come into agreement
with the SM expectation, we focus the discussion on possible NP
contributions to \BtoDsttaunuGen, where the discrepancy between
theory and experiment has recently increased.

The most thorough interpretation of a \BtoDsttaunuGen\ measurement in terms
of NP constraints was conducted by \babar\ for their
results\cite{ref:bbr-dtaunu-2012,ref:bbr-dtaunu-2013}. 
Within a type-II 2HDM (Eq.~(\ref{eq:R-NP-II})), they extracted
$\tan\beta/m_{H^+} = 0.44 \pm 0.02~\gev^-1$ from \RD\ and 
$\tan\beta/m_{H^+} = 0.75 \pm 0.04~\gev^-1$ from \RDst. From the
disagreement between these results, they ruled out the model with a
confidence level of at least $99.8\%$ for any value of
$\tan\beta/m_{H^+}$ (this includes the SM point of $\tan\beta/m_{H^+} =
0$, see Sec.~\ref{sec:dtaunu-babar-hadtag}), excluding a much broader
range of parameters than recent (albeit low-luminosity) 
LHC searches for a charged Higgs boson\cite{Aad:2012tj,Chatrchyan:2012vca}. 
A similar analysis has not been performed for the Belle
\BtoDsttaunuGen\ measurements. However, given the agreement of the
results of the two experiments, one can expect that combining their
results would yield even tighter limits on the parameter space.

\begin{figure}[!btph]
\centerline{
\includegraphics[width=0.7\textwidth]{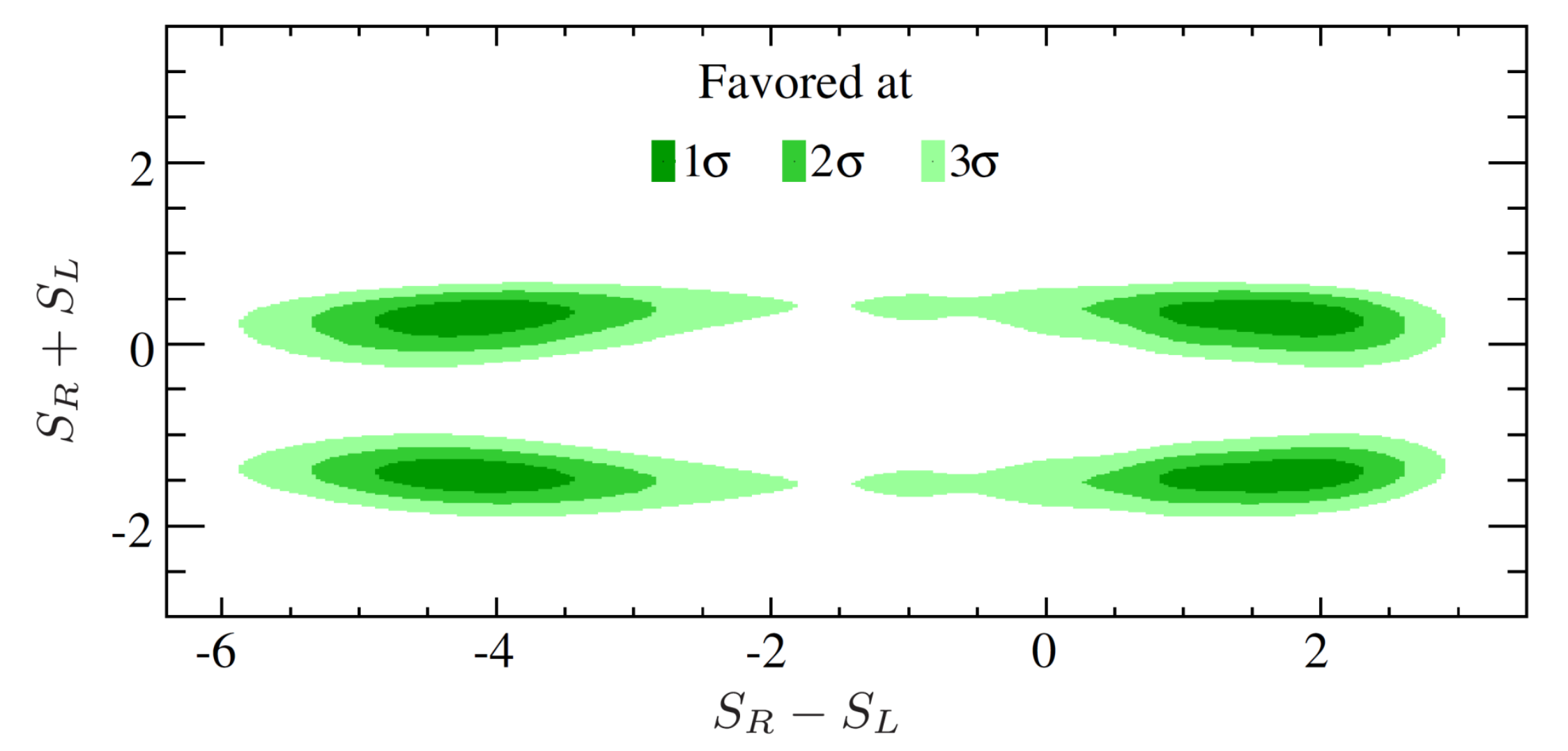}
}
\caption{Favored regions for real values of the type-III 2HDM
parameters $S_R$ and $S_L$ given the \babar\
results{\protect\cite{ref:bbr-dtaunu-2012,ref:bbr-dtaunu-2013}} for
\RDstGen.  Further analysis of the $q^2$ distributions disfavors
the two regions at $\Re(S_R + S_L) \sim -1.5$ with a
significance of at least $2.9\sigma$.
\protect\label{fig:Dtaunu-type3-constraints} }
\end{figure}

\babar\ also analyzed their \RDstGen\ results in the context of a
type-III 2HDM, restricting the analysis to real values of the parameters
$S_R$ and $S_L$ of Eq.~(\ref{eq:R-NP-III}). They found four favored
regions in the two-dimensional plane, shown in
Fig.~\ref{fig:Dtaunu-type3-constraints}.
Additional constraints were obtained by considering the $q^2$
distributions, in particular for \BtoDtaunu, which tends to shift to
higher values in the presence of a scalar contribution. For the two
favored regions shown in Fig.~\ref{fig:Dtaunu-type3-constraints} at
$\Re(S_R + S_L) \sim -1.5$, the expected $q^2$ distribution is
significantly harder than the spectrum measured in the data. As a
result, these regions were excluded with a significance of at least
$2.9\sigma$. Thus, only the two regions at $\Re(S_R + S_L) \sim 0.4$
were favored by the measurement. However, \babar\ noted that the $q^2$
spectra of the data were in better agreement with the SM than with
these regions, or with with 2HDM solutions with complex values of
$S_R$ and $S_L$.

\begin{figure}[!btph]
\addtocounter{footnote}{-1}
\centerline{
\includegraphics[width=0.9\textwidth]{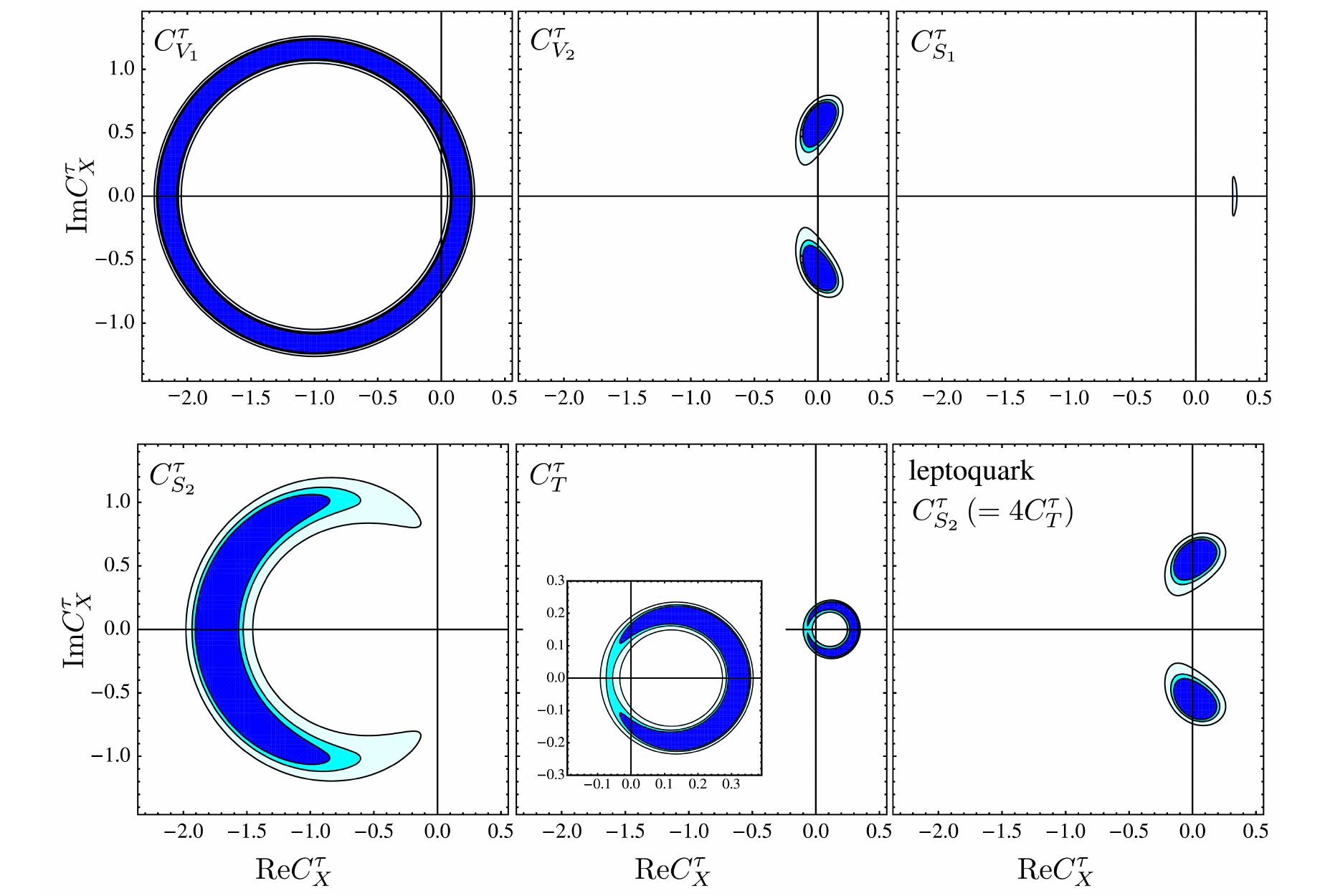}
}
\caption{90\% (light blue), 95\% (cyan), and 99\% (dark blue) 
confidence-level favored regions for the
  coefficients of Eq.~(\protect\ref{eq:dtaunu-H}), assuming
  that only one coefficient is non-zero{\protect\cite{Tanaka:2012nw}}.
  The conversion to the notation of Eq.~(\protect\ref{eq:dtaunu-H}) is
  $C^\tau_{V_1}=V_L$, 
  $C^\tau_{V_2}=V_R$, 
  $C^\tau_{S_1}=S_R$, 
  $C^\tau_{S_2}=S_L$, 
  $C^\tau_{T}=T_L$. 
  The bottom-right panel corresponds to a scalar leptoquark scenario, where
  $C^\tau_{S_2} = 4 C^\tau_T$.
  {\protect\label{fig:Tanaka-constraints}} }
\end{figure}

A number of authors have analyzed the \RDstGen\ results in terms of NP
contributions.
As an example, we quote some of the results of Tanaka and
Watanabe\cite{Tanaka:2012nw}, which were based on their combination of
the \babar\ and Belle results, $\RD = 0.305 \pm 0.012$, $\RDst = 0.252
\pm 0.004$.  In this model-independent analysis, they used the full
effective Hamiltonian of Eq.~(\ref{eq:dtaunu-H}). The
constraints they extracted on the coefficients
$V_{L,R}$, $S_{L,R}$, and $T_L$ are shown in
Fig.~\ref{fig:Tanaka-constraints}, under the assumption that only one
of the coefficients is non-zero.
Model-independent constraints allowing more than one coefficient to
vary at a time\cite{Dutta:2013qaa} or focusing on the tensor
operator\cite{Biancofiore:2013ki} have also been calculated.
Additional constraints have been determined for specific models, including
leptoquark
scenarios\cite{Fajfer:2012jt,Dorsner:2013tla,Sakaki:2013bfa}, chiral
$U(1)'$ models\cite{Ko:2012sv}, R-parity
violation\cite{Deshpande:2012rr}, sterile
neutrinos\cite{Abada:2013aba} , and nonuniversal left-right
models\cite{He:2012zp}.

\footnotetext{Fig.~{\protect\ref{fig:Tanaka-constraints}} is reprinted
  with permission from M.~Tanaka and R.~Watanabe,
  Phys.\ Rev.\ D. {\bf 87}, 0340281 (2013). Copyright (2013) by
  the American Physical Society.}

\section{Conclusions and Outlook}
\label{sec:conc}

Measurements of the branching fractions of the decays \Btotaunu\ and
\BtoDsttaunuGen\ have long been in some tension with the SM
expectations.  Recent measurements by \babar\ and Belle have
essentially removed the tension\cite{ref:bll-taunu-2013} in \Btotaunu\ and
increased it in \BtoDsttaunuGen\ to a level of at least
$3.2\sigma$\cite{ref:bbr-dtaunu-2012,ref:bbr-dtaunu-2013} and perhaps
as much as $4.8\sigma$\cite{ref:bozek-fpcp}.  The
\BtoDsttaunuGen\ results disfavor\cite{ref:bbr-dtaunu-2012} type-II
two-Higgs-doublet models with a confidence level of at least 99.8\%.
Constraints have been calculated on the parameter spaces of other
models and on Wilson coefficients within model-independent analyses.

\babar\ and Belle have used their full data
sets and best $B$-tagging methods for the \Btotaunu\ measurements.
Although further improvements in hadronic $B$ tagging
(Sec.~\ref{sec:hadronic-tag}) would yield some improvement in
efficiency, the overall sensitivity will not increase
substantially. Therefore, the current agreement of the world average for
$\br(\Btotaunu)$ with the SM prediction is expected to persist until the next
generation of $B$-factory experiments.

The situation is different for \BtoDsttaunuGen. Belle has yet to
perform the measurement of \RDstGen\ with their improved $B$-tagging
method. It remains to be seen whether this will bring the world
average into agreement with the SM or increase the overall discrepancy to
beyond 5~standard deviations.
\babar\ and Belle can also reconstruct \BtoDsttaunuGen\ with
semileptonic tagging, as well as obtain further insight into possible
new-physics contributions to this decay from an angular analysis
or from simpler measurements of forward-backward
asymmetries\cite{Celis:2012dk}.
The LHCb experiment may be able to contribute to the
\RDstGen\ measurements\cite{Kamenik:2008tj} with the $\tau$ lepton
identified in the decay $\tau^+\to\pi^+\pi^-\pi^+\bar\nu_\tau$, if the
large hadronic background can be suppressed down to a manageable
level.

Dramatic improvement in our understanding of \Btotaunu\ and
\BtoDsttaunuGen\ and of possible new-physics contributions will come
from the Belle-II experiment, which will have a data sample of about
$50\times 10^9$ $B\bar B$ pairs in the early-to-mid 2020's.  The expected
uncertainty on $\br(\Btotaunu)$ has been
estimated\cite{Meadows:2011bk} to be $4\times 10^{-6}$, a roughly
6-fold improvement over the current world average.
A similar improvement can be expected for \BtoDsttaunuGen. The precise
measurements of \RDstGen\ will settle the question of whether the
current tension with the SM is the result of a fluctuation or new
physics. In the latter case, measurements of the $q^2$ and angular
distributions are likely to have the precision needed for
differentiating between different new-physics models. Belle-II could
also measure $\br(B\to \bar D^{**}\tau^+\nu_\tau)$ for specific $\bar
D^{**}$ states, and have a dedicated run at the $\epem\to B_s \bar
B_s$ threshold to measure $\br(B_s\to D_s^{(*)-}\tau^+\nu_\tau)$.
These measurements would be by far less precise than those of
$\br(\BtoDsttaunuGen)$, but may turn out to shed light on the roles of
both new physics and hadronic processes in $\bar b \to \bar c \tau^+
\nu_\tau$ decays.

\section*{Acknowledgments}

I thank David Jaffe and Damir Becirevic for comments on the manuscript,
Dana Lindemann for technical assistance, and Andrzej Bozek for clarifications.

\end{document}